\documentclass[twocolumn,showpacs,floatfix,amsmath,amssymb,prb]{revtex4}
\usepackage[dvips]{graphicx}
\usepackage{latexsym}
\usepackage{dcolumn}
\usepackage{epsfig}
\usepackage{bm}
\usepackage{times,psfrag,subfigure}
\usepackage{dcolumn}% Align table columns on decimal point
\newcommand{\be}{\begin{equation}}
\newcommand{\ee}{\end{equation}}
\newcommand{\ba}{\begin{eqnarray}}
\newcommand{\ea}{\end{eqnarray}}
\newcommand{\baa}{\begin{eqnarray*}}
\newcommand{\eaa}{\end{eqnarray*}}

\begin{document}
        
\title{\textbf{Ground states of a frustrated
spin-$\frac{1}{2}$ antiferromagnet: Cs$_2$CuCl$_4$ in a magnetic field}}
\author{ 
M.\ Y. Veillette${}^1$, J.\ T. Chalker${}^1$ and R. Coldea${}^2$}
\affiliation{${}^1$ Theoretical Physics, University of Oxford, 1, Keble Road,
Oxford, OX1 3NP, United Kingdom\\}
\affiliation{${}^2$ Oxford Physics, Clarendon Laboratory, Parks Road, Oxford
OX1 3PU, United Kingdom\\}
%affiliation{$^2$ Oxford Physics, Clarendon Laboratory, Parks Road, Oxford, 0X1 3PU, United Kingdom\\}
\date{\today}
% 75.10.Jm Quantized spin models
% 75.45.+j Macroscopic quantum phenomena in magnetic systems
% 75.40.Gb Dynamic properties (dynamic susceptibility, spin waves,
%      spin diffusion, dynamic scaling, etc.)
% 73.43.Fj Novel experimental methods; measurements
% 75.30.Cr Saturation moments and magnetic susceptibilities
% 75.25.+z Spin arrangements in magnetically ordered materials
%       (including neutron and spin-polarized electron studies,
%       synchrotron-source x-ray scattering, etc.)
% 75.30.Ds Spin waves (for spin-wave resonance, see 76.50)
% 74.      Superconductivity
% 74.20.Mn Nonconventional mechanisms (spin fluctuations, polarons
%      and bipolarons, resonating valence bond model, anyon
%      mechanism, marginal Fermi liquid, Luttinger liquid, etc.)
% 05.30.Pr Fractional statistics systems (anyons, etc.)

\begin{abstract}
We present detailed calculations of the magnetic ground state properties
of Cs$_2$CuCl$_4$ in an applied magnetic field, and compare our results
with recent experiments. The material is described by a spin Hamiltonian, determined with 
precision in high field measurements, in which the main interaction is 
antiferromagnetic Heisenberg exchange between neighboring spins
on an anisotropic triangular lattice. An additional, weak
Dzyaloshinkii-Moriya interaction introduces easy-plane anisotropy, so that
behavior is different for transverse and longitudinal field
directions. We determine the phase diagram as a function of field strength
for both field directions at zero temperature, 
using a classical approximation as a first step. 
Building on this, we calculate the effect of quantum fluctuations on the 
ordering wavevector and components of the ordered moments, using both linear 
spinwave theory and a mapping to a Bose gas which gives exact results when 
the magnetization is almost saturated. Many aspects of the experimental data
are well accounted for by this approach.
\end{abstract}

\maketitle

%\draft

%\twocolumn
%\hsize\textwidth\columnwidth\hsize\csname @twocolumnfalse\endcsname
\section{Introduction}
\label{Introduction}

The layered, insulating magnet Cs$_{2}$CuCl$_{4}$ has attracted
intense recent experimental and theoretical 
attention.\cite{Coldea0,Coldea1,Coldea2,Coldea3,Coldea4, Bocquet,McKenzie1,Kim1,Zhang1,Zhang2,Wen1} 
Much of the interest arises because the material is a 
rare example of a spin $S=\frac{1}{2}$, triangular lattice
antiferromagnet.\cite{Coldea1,Coldea2,Coldea3}
Its small spin, quasi-two dimensionality and
geometric frustration are all features expected to enhance zero-point fluctuations in
N\'eel-ordered states, and to promote spin-liquid states.\cite{Anderson1, Anderson2, Review} 
Indeed, inelastic neutron scattering experiments on Cs$_{2}$CuCl$_{4}$ \cite{Coldea3}
have revealed extended scattering continua in the
dynamic structure factor, and various spin liquid
states\cite{McKenzie1,Wen1,Kim1} have
been proposed to explain this observation in terms of
fractionalized excitations. 
Nevertheless, at sufficiently low temperature Cs$_{2}$CuCl$_{4}$
displays conventional, magnetically ordered states over much of the phase
diagram spanned by magnetic field strength and 
direction.\cite{Coldea0,Coldea1,Coldea2,Coldea3}
In this paper, we develop a theoretical treatment of these ordered states
using two approaches. Starting from a classical description,
we discuss fluctuations using linear spinwave theory. And
starting from the fully polarized spin state reached at high field, we
discuss fluctuations as a dilute Bose gas of spin flips.
We present a detailed comparison of our results with experiment.

% Here, we summarized the experimental findings on Cs$_2$CuCl$_4$.

The low-temperature states of 
Cs$_2$CuCl$_4$ have been examined as a function of magnetic field
strength and direction, using neutron diffraction.\cite{Coldea1,Coldea2} At zero 
field, long-range order, in the form of an
incommensurate spiral spin structure, occurs below a N\'eel temperature of $T_N =
0.62$ K. The magnetic moments lie in an easy plane due to anisotropy arising from a 
Dzyaloshinskii-Moriya interaction. The presence of this interaction,
breaking SU(2) symmetry in the spin Hamiltonian, has profound consequences for the
behavior of the system in a magnetic field, and the ordering observed depends
on the field direction.\cite{Coldea1} 
Two field directions have been studied: {\it transverse} to the easy plane
(along the crystallographic $a$ direction) and {\it longitudinal} -- within the
easy plane (the crystallographic $b$-$c$ plane).
In a transverse field, spins cant out of the easy plane towards the field direction,
gaining Zeeman energy. Below a critical field of $B^a_{cr}=8.44$ T, 
ordered moments at different sites lie on a cone around the field direction. As the
critical field is approached the cone angle closes to zero, and above it
the magnetization is saturated. 
The behavior for a longitudinal
magnetic field is considerably more complex.\cite{Coldea1,Coldea4} For fields along the $c$
axis of strength $B^c$,
at weak fields, $B^c<1.4$ T, the anisotropy confines the spins in the
$b$-$c$ plane, creating a distorted cycloid.
In the field range $1.4 {\rm T} < B^c < 2.1 {\rm T}$, a second incommensurately
ordered phase appears.
% (see Fig. \ref{Lattice}, for geometry). 
At intermediate field strengths, in the range
$2.1$ T $<B^c<7.1$ T, no magnetic Bragg peaks have so far been reported. 
In stronger fields, magnetic Bragg peaks\cite{Coldea4}
at incommensurate wavevector are found up to the critical
field $B^c_{cr}=8.0$ T, beyond which the magnetization is 
saturated at low temperature. 

In this article our starting point is the spin Hamiltonian for
Cs$_2$CuCl$_4$, as
determined by high-field experiments.\cite{Coldea2} 
We discuss the symmetry of this Hamiltonian
and establish its ground-state phase diagram in transverse and longitudinal fields,
within a classical approximation. We find 
incommensurate phases of three types. 
Extending our treatment to include quantum fluctuations, we proceed 
in two ways. First, we set out linear spinwave theory, treating fluctuations
around the classical state at leading order using a standard $1/S$ expansion.
Although the expansion parameter is not small in the case at hand, results known 
for the nearest neighbor Heisenberg antiferromagnet on the square
\cite{Canali1, Igarashi1} and isotropic triangular \cite{Chubukov1} lattices 
suggest that linear spinwave theory is likely to be quite
accurate even for $S=\frac{1}{2}$. 
Second, supplementing the $1/S$ expansion, we apply theory for
a dilute Bose gas to spin flips in a system with almost saturated
magnetization. Using both methods, we determine quantum corrections to the
ordering wavevector and components of the local ordered moments
as a function of field strength. The results
depend markedly on the presence of a Dzyaloshinskii-Moriya interaction and on the
orientation of the applied magnetic field. We also investigate the effect of 
interlayer exchange, focussing on its influence on magnetic order
in a transverse field.
We compare our results extensively with experimental data.
 
The remainder of the paper is organized as follows. We introduce the spin Hamiltonian 
and discuss its symmetries in Sec. \ref{Spin Hamiltonian}. 
In Sec. \ref{Classical Analysis}, we develop
classical theory and establish the phase diagram within a classical approximation, in transverse and
longitudinal fields. We examine the effects of quantum fluctuations 
using the $1/S$ expansion and dilute Bose gas methods
in Sec. \ref{Linear Spin-wave Analysis}, calculating  static properties and comparing these with
experimental data. In Sec. \ref{Eccentricity} we consider interlayer coupling.
Finally, in Sec. \ref{Conclusion}, we summarize our conclusions.

\section{Crystal Structure and Spin Hamiltonian}
\label{Spin Hamiltonian}

\begin{figure}[ht]
\begin{center}
\includegraphics[width=8cm]{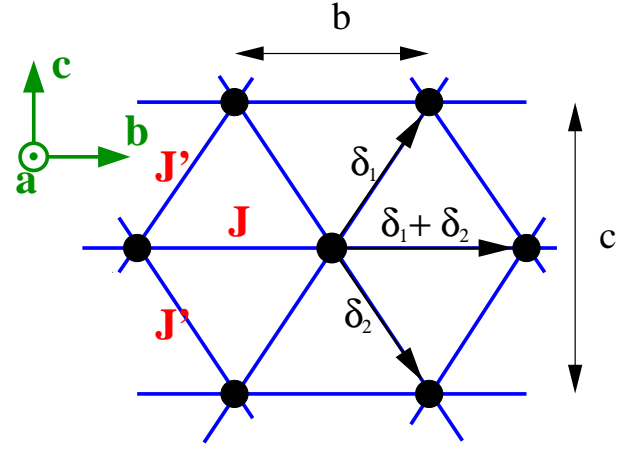} 
%\vspace{0.2cm}
\caption{Magnetic sites and exchange couplings within a single
layer of Cs$_{2}$CuCl$_{4}$.
Layers are stacked along the $a$-direction, with
interlayer spacing ${a}/{2}$ and a relative displacement in the $c$-direction.}
\label{Lattice}
\end{center}
\end{figure}

The magnetic moments in Cs$_{2}$CuCl$_{4}$ are carried by Cu$^{2+}$
ions. The orthorhombic unit cell contains four CuCl$^{2-}_4$
tetrahedra arranged in two layers in the $b$-$c$ plane.\cite{Coldea3}
The location of magnetic sites within a single layer is
illustrated in Fig.~\ref{Lattice}.
Exchange interactions are sufficiently weak that it is possible 
using laboratory magnetic fields to fully polarize
the moments at low temperature,  and the spin Hamiltonian has been determined
from a study of the excitation spectrum in a saturating transverse field.\cite{Coldea2}
This method has the advantage of yielding interaction constants
with the minimum of theoretical assumptions, since it focusses on the dynamics
of single spin flips. In this way it has been established that the
largest interaction is antiferromagnetic exchange $J$, coupling
neighboring spins along the chains, and that neighbors on adjacent
chains have a weaker exchange coupling $J^{\prime}$. 
In addition, the measurements indicate a
Dzyaloshinskii-Moriya (DM) exchange between chains, allowed by symmetry,\cite{Moriya60}
and a weak antiferromagnetic nearest-neighbor
interlayer coupling $J^{\prime \prime}$, which stabilizes long-range
magnetic order against thermal fluctuations.

%%%%%%%%%%%%%%%%%%%%%%%%%%%%%%%%%%%%%%%%%%%%%%%%%%%%%%%%%%%%%%%%%%%%%%%%%%
%%%%%%%%%%%%%%%%%%%%%%%%%%%%%%%%%%%%%%%%%%%%%%%%%%%%%%%%%%%%%%%%%%%%%%%% 

The model Hamiltonian, with experimentally determined parameter values
given in Table \ref{Hamiltonian Parameters}, is

%====================================================================
\begin{equation}
\mathcal{H}=\mathcal{H}_{0} + \mathcal{H}_{DM}+ \mathcal{H}_{B}\,,
\label{Hamiltonian00}
\end{equation}
%====================================================================
where $\mathcal{H}_0$ is the Heisenberg exchange energy,
$\mathcal{H}_{DM}$ represents the DM interaction, and
$\mathcal{H}_{B}$ is the Zeeman energy in an applied magnetic
field. Denoting spin-$\frac{1}{2}$ operators at the sites ${\bf R}$
of a stacked anisotropic triangular lattice by ${\bf S}_{\bf R}$,
the exchange energy is
%====================================================================
%\begin{equation}
%\mathcal{H}_0=\frac{1}{2}\sum_{{\bf R},{\bf\delta }}J_{\bf\delta}{\bf S}_{\bf R}\cdot {\bf S}_{\bf{R}+\bm{\delta}},
%\label{Hamiltonian0}
%\end{equation}
%================================================================
%====================================================================
\begin{eqnarray}
\mathcal{H}_0=\sum_{{\bf R}} &&\big[J 
{\bf S}_{\bf R}\cdot {\bf
S}_{\bf{R}+\bm{\delta_{1}+{\delta_{2}}}}\nonumber \\
\phantom{\sum_{{\bf R}}}&&+J^{\prime} \left( {\bf S}_{\bf R}\cdot {\bf
S}_{\bf{R}+\bm{\delta_{1}}} + {\bf S}_{\bf R}\cdot {\bf
S}_{\bf{R}+\bm{\delta_{2}}} \right)\nonumber \\
\phantom{\sum_{{\bf R}}}&&+J^{\prime \prime} {\bf S}_{\bf R}\cdot {\bf
S}_{\bf{R}+\bm{\delta_{3}}}\big]\,,
\label{Hamiltonian0}
\end{eqnarray}
%================================================================
%where ${\bm \delta}$ is a vector between nearest-neighbor sites. 
where the nearest neighbor vectors ${\bm \delta_{1}}$ and ${\bm \delta_{2}}$
are indicated in Fig. \ref{Lattice} and the out-of-plane vector
${\bm \delta_{3}}$ connects spins on adjacent layers.
The DM energy is
%=================================================================
\begin{equation}
\mathcal{H}_{DM}\!\!=-\sum_{\bf R} (-1)^n {\bf
D}\cdot {\bf S_{\bf R}}\times \lbrack {\bf
S_{\bf{R}+\bm{\delta}_{1}}}+ {\bf
S_{\bf{R}+\bm{\delta}_{2}}}
\rbrack\,,
\label{hamiltonianDM}
\end{equation}
%=================================================================
where ${\bf D}=(D,0,0)$ is a vector associated with the oriented bond
between the two coupled spins and  $n$
is a layer index. The factor $(-1)^n$ indicates that 
the %direction of the vector ${\bf D}$
interaction alternates between even and odd layers, which
are inverted version of one another.
% The nearest neighbor vector
%$\delta_{1}, \delta_{2}$ are indicated in Fig. \ref{Lattice} and
The Zeeman energy arising from a magnetic field ${\bf
B}=(B^a,B^b,B^c)$ is
%===============================================================
\begin{equation}
\mathcal{H}_{B}= - \sum_{R} g_i \mu_B B^{i}  S^{i}_{\bf{R}}\,,
\end{equation}
%=================================================================
where $g$ is the gyromagnetic tensor $g=(2.20, 2.08, 2.30)$.
\cite{Gfactor}

\begin{table}
\caption{Hamiltonian parameters, from Ref.~\onlinecite{Coldea2}.}
\label{Hamiltonian Parameters}
\begin{ruledtabular}
\begin{tabular}{cc}
Parameters & Exp. \\
\hline
$J$ (meV) & 0.374(5) \\ $J^{\prime}$ (meV) & 0.128(5) \\ $J^{\prime
\prime}$ (meV) & 0.017(2) \\ $D$ (meV) & 0.020(2) \\
\end{tabular}
\end{ruledtabular}
\end{table}

We omit the dipole-dipole interaction and several small effects, including
a relative offset of the Cu ions along $c$ between adjacent layers,
a small component of the ${\bf D}$ vector perpendicular to the $a$-axis
and possible anisotropy of the exchange interactions in spin space.

At the classical level, the intrachain coupling $J$ favors a staggered
magnetization in the spin chains and the interchain coupling $J^{\prime}$
frustrates this state. As $J^{\prime}/J$ is varied, ${\cal H}_0$ interpolates
between the fully frustrated Hamiltonian  for the isotropic triangular
lattice ($J^{\prime}=J$) and that for uncoupled
one dimensional spin chains $(J^{\prime}=0)$. The DM
interaction favors states in which spins lie in the $b$-$c$ plane, with a
rotation of $\pi/2$ between adjacent spin chains.

It is convenient at this point to introduce notation associated with reciprocal space.
We express wavevectors in terms of the reciprocal lattice vectors,
writing ${\bf Q}=(h,k,l)$ as shorthand for $2\pi(h/a,k/b,l/c)$.
The Fourier transforms of the exchange and DM interactions are
\be
J_{\bf Q}= J \cos (2\pi k )+ 2 J^{\prime} \cos ( \pi k) \cos ( \pi
l)
\ee 
and
\be
D_{\bf Q}= -2D \sin (\pi k) \cos( \pi l )\,.
\ee
When considering transverse magnetic fields, these appear in the combination
\be
J^T_{\bf Q}= J_{\bf Q}+ D_{\bf Q}\,.
\ee

%\subsection{Symmetry}
%\label{Symmetry}
We close this section with a discussion of the symmetry of the
Hamiltonian $\cal H$. While 
$\mathcal{H}_0$ has full SU(2) spin symmetry, $\mathcal{H}_{DM}$
has a lower, ${\mathbb Z}_2 \otimes {\rm U(1)}$ symmetry. Here, U(1)
arises from spin rotations around the ${\bf D}$ vector, and ${\mathbb Z}_2$
originates from invariance under the combination of 
space inversion (${\bf R}
\rightarrow -{\bf R}$) and the spin operations 
%$S^a \rightarrow - S^a$,
%$S^b \rightarrow S^b$, $S^c \rightarrow -S^c$. 
\ba
{\bf S} \times \hat{\bf x} & \rightarrow & -{\bf S} \times \hat{\bf x}
\nonumber \\
{\bf S} \cdot \hat{\bf x} & \rightarrow & {\bf S} \cdot \hat{\bf x}
\ea
where $\hat{\bf x}$ is an arbitrary unit vector in the $b$-$c$
plane.  To illustrate the nature
of the ${\mathbb Z}_2$ symmetry, one can consider the chiral scalar $K=
\sum_{\bigtriangleup} {\bf S}_{{\bf 1}} \cdot \left( {\bf S}_{\bf 2}
\times {\bf S}_{\bf 3} \right)$, where the spin product is performed in
a cyclical fashion over all triangular plaquettes. Under the ${\mathbb Z}_2$ operation,
$\mathcal{H}_0 + \mathcal{H}_{DM}$ is invariant but $K
\rightarrow -K$.
% indicating that the chiral scalar is can be used as an
%order parameter to characterize the two states. 
The inclusion of $\mathcal{H}_B$ further reduces the symmetry, to U(1) in a
transverse magnetic field (with $S^a$ a conserved quantity), 
and to ${\mathbb Z}_2$ in a longitudinal field.

\section{Classical Analysis}
\label{Classical Analysis}

The classical approximation consists of treating the spin operators
${\bf S}$ as classical vectors of length $S=1/2$. The Hamiltonian
then becomes an energy functional 
%$E_{cl}({\bf S})$ 
which can be minimized to determine the magnetic structure. 
%In zero magnetic field, the $b-c$ plane defines an easy-plane. 
Omitting interlayer exchange and DM interactions,
%Restricting the energy functional to a single layer and neglecting the DM term, 
the classical ground state in zero field is a spin spiral
%====================================================================
\be
{\bf S}_{{\bf R}}= S
\begin{pmatrix}
 0 \\
\cos  (  {\bf Q}^{\star}_{cl} \cdot {\bf R} +\alpha )\\
\sin  (  {\bf Q}^{\star}_{cl} \cdot {\bf R} + \alpha )
\end{pmatrix},
\ee
%=============================================
where the arbitrary phase $\alpha$ reflects spontaneous
breaking of the U(1) symmetry and the wavevector ${\bf Q}^{\star}_{cl}$ is determined by 
minimizing the exchange energy $J_{\bf Q}$.
%\be
%J_{\bf Q}= J \cos (2\pi k )+ 2 J^{\prime} \cos ( \pi k) \cos ( \pi l).
%\ee 
%Note that throughout this paper, we express the wavevectors ${\bf
%Q}=(h,k,l)$ in reciprocal lattice units of $( \frac{2 \pi}{a}, \frac{2
%\pi}{b}, \frac{2\pi}{c})$.  
We find ${\bf Q}^{\star}_{cl}= \pm
(0,1/2+\epsilon^\star_{cl},0)$ where  $\epsilon^{\star}_{cl} = \pi^{-1}
 \arcsin ({J^{\prime}}/{2 J})=0.0547$.  

With ${\mathcal H}_{DM}$ included, the degeneracy of the ground state 
with respect to the sign of
the ordering wavevector is broken. Since the sign of the DM term alternates on adjacent layers, the
direction of the wavevector alternates from layer to layer to give the
spin structure (setting $\alpha=0$)
%====================================================================
\be
{\bf S}_{{\bf R}}= S
\begin{pmatrix}
 0\\
 \cos  ( {\bf Q}_{cl} \cdot {\bf R}) \\
(-1)^n  \sin ( {\bf Q}_{cl} \cdot {\bf R} )
\end{pmatrix},
\label{spinstructureinzerofield}
\ee
%=============================================
where now ${\bf Q}_{cl}$ 
is determined by the minimum of $J^T_{\bf Q}$.
%\be
%J^T_{\bf Q}= J_{\bf Q}+ D_{\bf Q},
%\ee
%where  
%\be
%D_{\bf Q}= -2D \sin (\pi k) \cos( \pi l )\,.
%\ee
We find ${\bf Q}_{cl}=(0,1/2+\epsilon_{cl},0)$ with
$\epsilon_{cl}=0.0533$. 
%The ground state describes the solution of two cycloids that counter rotate on even and odd layers. 

The classical ground state in the presence of a transverse magnetic
field can be found easily because U(1) symmetry ensures that
only Fourier components with ${\bf Q} = {\bf 0}$ and ${\bf Q} = {\bf Q}_{cl}$ 
contribute to the spin configuration. 
The spiral order of spin components within the $b$-$c$ plane
is preserved, and the spins cant towards the field direction to produce a cone
state with
%====================================================================
\be
{\bf S}_{{\bf R}}= S
\begin{pmatrix} 
\sin \theta_{0} \\
\cos \theta_{0} \cos  ( {\bf Q}_{cl} \cdot {\bf R} )\\
(-1)^n \cos \theta_{0} \sin ( {\bf Q}_{cl} \cdot {\bf R})
\end{pmatrix},
\label{conestate}
\ee
%====================================================================
where, measuring magnetic field in the reduced units $h^i=g^i \mu_b B^i/S$,
$\sin \theta_{0}= h^a/h^a_{cr}$.  The critical field in
reduced units is $h^a_{cr}= 2 \left[ J^{T}_{\bf 0}- J^{T}_{\bf Q}
\right]$, giving $B^a_{cr}=8.36$ T. The same expression
for  $B^a_{cr}$ also emerges from an exact treatment of the
quantum Hamiltonian for a single layer (see Ref.~\onlinecite{Coldea2} and 
Sec.~\ref{Dilute Bose Gas});
the small difference between this value and the experimental one\cite{Coldea2} of
$8.44$ T is partly due to interlayer exchange (see Sec.~\ref{Eccentricity}). 
At higher fields, the spins are fully polarized along the field
direction.

Ground states in a longitudinal field are considerably more complex
because the magnetic field breaks U(1) symmetry and many Fourier
harmonics contribute to the spin configuration.
% At low fields
%the anisotropy of the DM interaction confines the
%spins to the $b$-$c$ plane and spin directions within this plane
%are determined by competition between exchange and Zeeman energy,
%resulting in a distorted cycloidal structure.
A useful guide to the behavior one should expect is provided by 
results for
frustrated magnetic systems in a magnetic field, with single ion anisotropy rather than DM interactions.
In that case, if anisotropy is weak, there is a first order transition 
between a distorted cycloid state at low field, in which spins are confined to
the easy plane, and an incommensurate cone structure with its axis
along the field direction at high field.\cite{Nagamiya0}

To investigate such phenomena in the problem we are
concerned with, we have studied spin configurations obtained by minimizing
the classical energy functional numerically. We use periodic
boundary conditions with a period of over 1000 sites in the $b$-direction,
and have examined many minima for a range of values of longitudinal
fields.
We find two phases separated by a first order
transition. For ${h}/{h_{cr}} < 0.35$, the zero-field
spin spiral evolves smoothly into a distorted cycloid in which spins 
lie in the $b$-$c$ plane. This state has a continuous degeneracy
associated with phason modes.\cite{Elliot}
For fields in the range $ 0.35 < {h}/{h_{cr}} < 1$, an
incommensurate out-of-plane solution is optimal. It has
a non-zero value for the chiral scalar $K$ and therefore breaks ${\mathbb Z}_2$
symmetry. A very good approximation to the out-of-plane numerical solution is provided
by the expression
\be
{\bf S}_{{\bf R}}= S
\begin{pmatrix}
\cos \theta_{0} \cos ( {\bf Q} \cdot {\bf R})  \cos\eta+ \sin\theta_{0} \sin \eta \\ 
(-1)^n \cos \theta_{0} \sin ({\bf Q} \cdot {\bf R}) \\ 
\sin\theta_{0} \cos\eta- \cos \theta_{0} \cos ({\bf Q} \cdot {\bf R})  \sin\eta 
\end{pmatrix}.
\label{tilted-cone}
\ee
In this approximation, only the Fourier components ${\bf 0}$ and ${\bf Q}$ appear,
and the ordering wavevector is
within a few percent of $ {\bf Q}^{\star}_{cl}$. Spin directions at different
sites form a cone, which has a height $S \sin\theta_{0}$ and an axis
lying in the $a$-$c$ plane, tilted at an angle $\eta$ to the $c$-direction.
Moving from site to site in the $b$-direction, 
the spin projection onto the easy plane traces out an
ellipse. The eccentricity of this ellipse is associated with 
a non-zero DM energy, and $\eta \propto D$ for small $D$.
A second ground state, related by ${\mathbb Z}_2$ symmetry to the first, is generated by 
the operation: ${\bf Q} \rightarrow -{\bf
Q}$ and $\eta \rightarrow -\eta $.

In spite of the proximity of the incommensurate wavevector to the
commensurate value $(0,\frac{1}{2},0)$, the commensurate
states are found to be well separated in energy from the
incommensurate solutions, within a classical treatment.

\begin{figure}[ht]
\begin{center}
\includegraphics[angle=0,width=8cm]{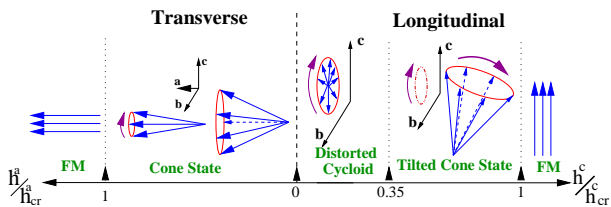} 
%\vspace{0.2cm}
\caption{Phase diagram in the classical limit, with a schematic representation of the different phases.
Transitions between the cone states and the ferromagnetic states are second order. 
The distorted cycloid and the tilted cone states are separated by a first order transition.}
\label{Phase Diagram}
\end{center}
\end{figure}

The results of this classical analysis are summarized in 
Fig.~\ref{Phase Diagram}. Behavior in a transverse field is in qualitative
agreement with the experimental findings outlined in Sec.~\ref{Introduction}.
We delay a quantitative comparison between theory and experiment until
after our discussion of the effects of quantum fluctuations in Sec.~\ref{Linear Spin-wave Analysis}.
Contrastingly, observed behavior in a longitudinal field shows different
features from the classical phase diagram. In particular,
the state found in the field range $2.1 {\rm T} < B^c < 7.1 {\rm T}$
does not appear classically.
%%%%%%%%%%%%%%%%%%%%%%%%%%%%%%%%%%%%%%%%%%%%%%%%%%%%%%%%%%%%%%%%%%%%%%%%%%
%%%%%%%%%%%%%%%%%%%%%%%%%%%%%%%%%%%%%%%%%%%%%%%%%%%%%%%%%%%%%%%%%%%%%%%% 

\section{Quantum Fluctuations}
\label{Linear Spin-wave Analysis}

The classical ground states determined in Sec. \ref{Classical Analysis}
provide a starting point for a treatment of quantum fluctuations.
This can be approached using either a $1/S$ expansion, or directly
for $S=\frac{1}{2}$ by expanding in powers of the density of reversed
spins in a polarized background, viewing these as a
dilute Bose gas. While the $1/S$ expansion is uncontrolled when applied
to Cs$_{2}$CuCl$_{4}$, it is known to produce quite
accurate results for some simpler two-dimensional
$S=\frac{1}{2}$ systems.\cite{Canali1, Igarashi1,Chubukov1}
Conversely, the density of reversed spins is controlled by field strength
and the expansion parameter is $(1-h/h_{cr})$.
%for transverse fields, in which $U(1)$ symmetry ensures that their number is conserved,
%the expansion parameter is$(1-h^a/h^a_{cr})$, while for longitudinal fields
%it is the inverse field strength.
It is worth pointing out that quantum fluctuations in Cs$_2$CuCl$_4$ do not break a classical
degeneracy, as is the case for the isotropic triangular lattice antiferromagnets in a field,\cite{Golosov}
but are likely to have substantial quantitative effects on ground state properties.

\subsection{Large S-expansion}

We now turn to a description of the calculations. The procedure is
standard: starting from a classical, ordered state 
we use the Holstein-Primakoff transformation to obtain a bosonic
Hamiltonian.\cite{Nagamiya,White01,Shiba2,Shiba1,Shiba3,Zhitomirsky02,Zhitomirsky98}
Considering only the quadratic part of this Hamiltonian, we obtain
the leading quantum contribution in a $1/S$ expansion.

\subsubsection{Transverse Field}

In a transverse magnetic field, the classical ground state is an 
incommensurately ordered spin cone with wavevector ${\bf Q}$, given by Eq.~(\ref{conestate}). 
We introduce a rotating coordinate system in spin space, via
%\begin{widetext}
\ba
\begin{pmatrix}
S^a_{\bf R}\\ S^b_{\bf R}\\ S^c_{\bf R}
\end{pmatrix}
&=&
\begin{pmatrix}
1 & 0 &0 \\ 0 & \cos ({\bf Q} \cdot {\bf R}) & -(-)^n \sin ({\bf Q}
\cdot {\bf R}) \\ 0 & (-)^n \sin ({\bf Q} \cdot {\bf R}) & \cos ({\bf
Q} \cdot {\bf R})
\end{pmatrix}\times \nonumber \\
&& \qquad \qquad \times\begin{pmatrix}
\cos \theta & 0 & \sin \theta  \\
0 & 1 & 0 \\
-\sin \theta & 0 & \cos \theta  
\end{pmatrix}
\begin{pmatrix}
S^{x}_{\bf R}\\ S^{y}_{\bf R}\\ S^{z}_{\bf
R}
\end{pmatrix},
\ea
%\end{widetext}
chosen so that the $z$ axis at each site is aligned
with the classical spin direction. A central objective 
of this section is to calculate quantum corrections to 
classical values of 
the ordering wavevector ${\bf Q}$ and the canting angle $\theta$.
We omit the small interlayer exchange $J^{\prime \prime}$, postponing a
discussion of some of its effects to
Sec.~\ref{Eccentricity}. 

The Holstein-Primakoff transformation is
\ba
S^{x}_{\bf R}&=& \frac{ \sqrt{2S}}{2} \left (
\phi^{\dagger}_{\bf R} +\phi^{\phantom{\dagger}}_{\bf R} \right),\nonumber \\
%\qquad
 S^{y}_{\bf R}&=& i \frac{ \sqrt{2S}}{2} \left(
\phi^{\dagger}_{\bf R} -\phi^{\phantom{\dagger}}_{\bf R} \right),\nonumber \\
%\qquad 
S^{z}_{\bf R}&=& S - \phi^{\dagger}_{\bf R}
\phi^{\phantom{\dagger}}_{\bf R},
\ea
where the boson creation and annihilation
operators satisfy the commutation relation $\left
[ \phi^{\phantom{\dagger}}_{\bf R}, \phi^{\dagger}_{{\bf R}^\prime}
\right]= \delta_{\bf R, R^{\prime}}$. Introducing the Fourier
transform
\be
 \phi^{\dagger}_{{\bf k}}=\frac{1}{ \sqrt{N}} \sum_{\bf R} \phi^{\dagger}_{\bf R}
 e^{-i {\bf k} \cdot {\bf R}},
\ee
for a lattice of $N$ sites, the
Hamiltonian of Eq.~(\ref{Hamiltonian00}) takes the form
%==========================================================================
\be
{\mathcal H}= {\mathcal H}_0+{\mathcal H}_1+{\mathcal H}_2 + \cdots,
\ee
where ${\mathcal H}_{n}$ is proportional to $S^{2 -n/2}$
and consists of products of $n$ normal-ordered boson operators.  
%The odd terms of this expansion are present only for a non-collinear spin
%structure and therefore vanish in zero magnetic field. 
The leading terms are 
%===========================================================
\ba
%{\mathcal H}_0 &=& N S^2 \left( J^{T}_{\bf Q}- \frac{(h^a)^2}{4 \left
%[ J^{T}_{\bf 0}- J^{T}_{\bf Q} \right]} \right), \\ 
{\mathcal H}_0 &=& N S^2 \left( J^{T}_{\bf Q}+ \left[ J^T_{\bf 0}
-J^T_{\bf Q} \right] \sin^2 \theta -h^a \sin \theta \right) \nonumber \\
{\mathcal H}_1 &=& \sqrt{\frac{NS^3}{2}}  \cos \theta  \left( 2
\left[ J^{T}_{\bf 0}-J^T_{\bf Q} \right]  \sin \theta-h^a 
\right)  \left({ \phi^\dagger_{\bf 0} +
\phi^{\phantom{\dagger}}_{\bf 0 }}\right)  \nonumber\\
{\mathcal H}_2 &=& N S \left( J^{T}_{\bf Q} + \left[ J^T_{\bf
0}-J^T_{\bf Q} \right] \sin^2 \theta -\frac{h^a}{2} \sin \theta  \right)\nonumber\\
&+& S \sum_{\bf k} \left(A_{\bf k}+C_{\bf k}
\right) \left( \phi^\dagger_{\bf k} \phi^{\phantom{\dagger}}_{\bf
k}+\phi^{\phantom{\dagger}}_{\bf k} \phi^{{\dagger}}_{\bf k} \right)\nonumber\\ &+&
B_{\bf k} \left( \phi^\dagger_{\bf -k} \phi^{{\dagger}}_{\bf
k}+\phi^{\phantom{\dagger}}_{\bf -k} \phi^{\phantom{\dagger}}_{\bf k}
\right) 
\ea
%===========================================================
where the sum on ${\bf k}$ is performed over the first Brillouin
zone and
%===========================================================
\ba
 A_{\bf k}&=&  \frac{1}{4}\left[2J_{\bf k}+J^{T}_{\bf Q+k}+ J^{T}_{\bf Q-k} \right] - 
 J^{T}_{\bf Q} \nonumber \\ &-&
\frac{1}{4}\left[  2 J_{\bf k}-J^{T}_{\bf Q+k}- J^{T}_{\bf Q-k}
 \right] \sin^2 \theta \nonumber \\ &-& \left[ J^T_{\bf 0}-J^T_{\bf Q} \right]
 \sin^2 \theta + \frac{h^a}{2} \sin \theta \nonumber \\
 B_{\bf k}&=& \frac{1}{4}\left[ 2 J_{\bf k}-J^{T}_{\bf Q + k}- J^{T}_{\bf Q -
 k}\right] \cos^2 \theta \nonumber
 \\ C_{\bf k}&=& \frac{1}{2} \left[ J^{T}_{\bf Q + k}- J^{T}_{\bf Q - k}\right]
 \sin \theta.
\ea
%===========================================================
The coefficients $A_{\bf k}$ and $B_{\bf k}$ are even functions of
${\bf k}$, whereas $C_{\bf k}$ is an odd function of ${\bf k}$. 
The term ${\mathcal H}_{1}$  is linear in the bosonic operators
and vanishes if the canting angle $\theta$ takes its classical value, $\theta_{0}$.
The  quadratic Hamiltonian is diagonalized by the Bogoliubov transformation
\ba
\phi_{\bf k}&=& u_{\bf k} \gamma^{\phantom{\dagger}}_{\bf k} + v_{\bf
k} \gamma^{\dagger}_{\bf -k}, \nonumber \\
\phi^{\dagger}_{-\bf k}&=& v_{\bf k} \gamma^{\phantom{\dagger}}_{\bf k} + u_{\bf k}
\gamma^{\dagger}_{\bf -k}, 
\label{Bogoliubov}
\ea
where 
\ba
u^2_{\bf k}&=&1+v^2_{\bf k}= \frac{1}{2} \left( \frac{A_{\bf k}}{\sqrt
{ A^2_{\bf k} -B^2_{\bf k}}} +1 \right) \nonumber \\
% \qquad
u_{\bf k} v_{\bf k} &=&  \frac{1}{2} \frac{-B_{\bf k}}{\sqrt
{ A^2_{\bf k} -B^2_{\bf k}}}.\label{Bogoliubov2}
\ea
The diagonal form of the quadratic Hamiltonian is
%===========================================================
\ba
{\mathcal H}_2 &=& N S \left( J^{T}_{\bf Q} + \left[ J^T_{\bf
0}-J^T_{\bf Q} \right]\sin^2 \theta  - \frac{h^a}{2} \sin \theta
\right)\nonumber\\ &+& 2 S \sum_{\bf k}
\Omega_{\bf k} \left( \gamma^{\dagger}_{\bf k}
\gamma^{\phantom{\dagger}}_{\bf k} + \frac{1}{2} \right),
\ea
%===========================================================
where $\Omega_{\bf k}=\sqrt{ A^2_{\bf k} -B^2_{\bf k}}
+C^{\phantom{2}}_{\bf k}$ is the spinwave dispersion relation.\cite{White01, Nagamiya}
Setting $\theta=\theta_0$,
the spectrum has a gapless mode at ${\bf k}=0$ as a result of the U(1)
symmetry. The low-lying excitations are spin oscillations within the
plane of the cycloid. For an SU(2) symmetric Hamiltonian, a second
Goldstone mode is present at the ordering wavevector ${\bf Q}$ of
the cycloid. The low-lying excitations in this case involve
oscillations of the orientation the plane of the cycloid. For the
Hamiltonian we are concerned with, the DM interaction 
lifts the SU(2) symmetry and generates an excitation gap 
at wavevector ${\bf Q}$,
which becomes wider in an applied magnetic field.
Recently, it has been shown that the spin-wave spectrum 
of an antiferromagnet in a strong magnetic field is kinematically
unstable to two-magnon decay.\cite{Zhitomirsky01} Here 
we neglect such decay processes and retain only harmonic terms
in the Hamiltonian.
%The hybridization of the two-magnon continuum
%with the magnon branch leads to damping and the disappearance of the
%single-magnon branch in most of the Brillouin zone. In principle,
%these effects can be calculated systematically in higher order in
%$1/S$ but a strong damping casts doubts on the accuracy of the
%first-order $1/S$ calculation which is based on the single-magnon
%modes. However, we argue that the most significant contribution to the
%quantum fluctuation originates from low lying excitations. At low
%energy, the two-magnon decay processes have restricted allowed phase
%space due to kinematic constraints and therefore the weak mixing to
%the two magnon continuum preserve the character of these single
%particle excitations. The low damping associated with the low-energy
%modes leads us to conclude that it is appropriate to calculate the
%first-order contribution by summing over the single particle
%excitations.

The ground-state energy, 
omitting terms ${\cal O}(S^0)$ and higher, is then
%  A_{\bf k}&=& \left[ J^{T}_{\bf Q+k}+ J^{T}_{\bf Q-k} \right] \left( 1 + \left(\frac{h^a}{h^a_{cr}}\right)^2 \right)  + 2 J_{\bf k} \left( 1 - \left( \frac{h^a}{h^a_{cr}} \right)^2 \right)  -4 J^{T}_{\bf Q}  \\
%===========================================================
\ba
E\equiv \langle {\mathcal H} \rangle &=& N S (S+1) \left( J^{T}_{\bf Q} + \left[
 J^T_{\bf 0} - J^T_{\bf Q}\right] \sin^2 \theta \right)
 \nonumber\\ &-& N S(S+1/2)
 h^a \sin
 \theta + S \sum_{\bf k} \Omega_{\bf k}\,.
\label{RenormalizedQ}
\ea
%===========================================================
The ordering wavevector is to be determined by
minimizing $E$ with respect to ${\bf Q}$.
Following this procedure, the $1/S$ correction to $E$ comes not only from the
zero-point fluctuations but also from the renormalization of 
$\theta$ and ${\bf Q}$. 

\begin{figure}[ht]
\begin{center}
\begin{picture}(200,150)
\put(0,0){\epsfig{file=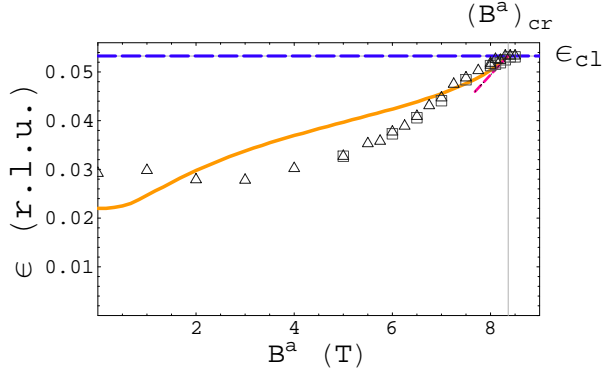,width=8cm}}
%\put(110,20){\epsfig{file=TransIncS1.eps,width=2.8cm}}
\end{picture}

%\vspace{0.2cm}
\caption{The incommensuration $\epsilon$ as a function of transverse magnetic
  field strength $B^a$. Full line: result from $1/S$ expansion.
  Long dashed line : result from the classical theory. 
Short dashed line: linear variation of $\epsilon$ with $B^a$, from calculation for
  dilute Bose gas of spin flips.
  Symbols are the experimental results taken from Fig 3c of
  Ref.~\onlinecite{Coldea2} taken at $T=0.20$K: $\triangle$ from magnetic Bragg
  peaks at ${\bf Q} = (0,1.5-\epsilon,0)$, and $\Box$ from peaks at
  ${\bf Q} = (0,0.5-\epsilon,1)$.
%  theoretical dimensionless magnetic field have been made
%  unit full using gyromagnetic factors determined by S. Bailleul {\em et al.}
%  \cite{Gfactor}.}
% In the inset, the same data is presented near the critical field.
}
\label{TransverseIncommensuration}
\end{center}
\end{figure}

Results for ${\bf Q}\equiv(0,1/2+\epsilon,0)$ to ${\cal O}(S^{-1})$ are shown in
Fig. \ref{TransverseIncommensuration}, together with data 
from Ref.~\onlinecite{Coldea2}. At the critical field
$h^a_{cr}$ we find, in agreement with the experiment, that ${\bf Q}$ takes the classical
value ${\bf Q}_{cl}$, which is field-independent.
This is a consequence of the fact that the ferromagnetically polarized state
is an exact eigenstate of the Hamiltonian with vanishing zero-point energy.
%as demonstrated explicitly to ${\cal O}(S^{-1})$ by the fact that
%$ J^T_{\bf Q}+ \sum_{k} \Omega_{\bf k}=0$ for $h^a=h^a_{cr}$ and $\sin(\theta)=1$.
At lower fields, fluctuations renormalize ${\bf Q}$, which
decreases with decreasing field: the zero-field value
of the incommensuration $\epsilon=0.21$ is significantly reduced from 
its value at the critical field. This reduction can be understood on the basis that 
zero-point energy in antiferromagnets generally is lowered 
for states with collinear spins. The states we are
concerned with are close to the collinear state with
${\bf Q} = (0,1/2,0)$, but have lower classical energies. 
With decreasing field, quantum
fluctuations are enhanced and drive the incommensurate
wavevector towards the commensurate value. As a technical aside,
we note that calculations are simplified by the presence of DM interactions,
since without them the Goldstone mode at wavevector ${\bf Q}$, which appears
as $h \rightarrow 0$, necessitates a self-consistent treatment of
quantum fluctuations.
%%%%%%%%%%%%%%%%%%%%%%%%%%%%%%%%%%%%%%%%%%%%%%%%%%%%%%%%%%%%%%%%%%%%%%%%%%
%%%%%%%%%%%%%%%%%%%%%%%%%%%%%%%%%%%%%%%%%%%%%%%%%%%%%%%%%%%%%%%%%%%%%%%% 

\begin{figure}[h2t]
\centerline{\begin{picture}(200,150)
\put(-20,0){\includegraphics[width=8cm]{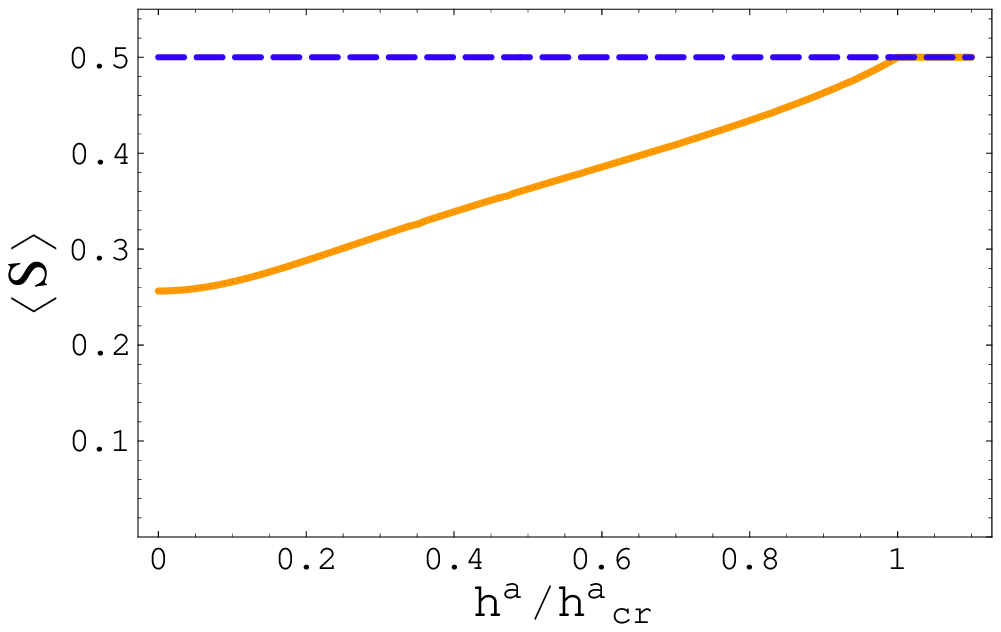}}
\put(85,25){\includegraphics[width=4cm]{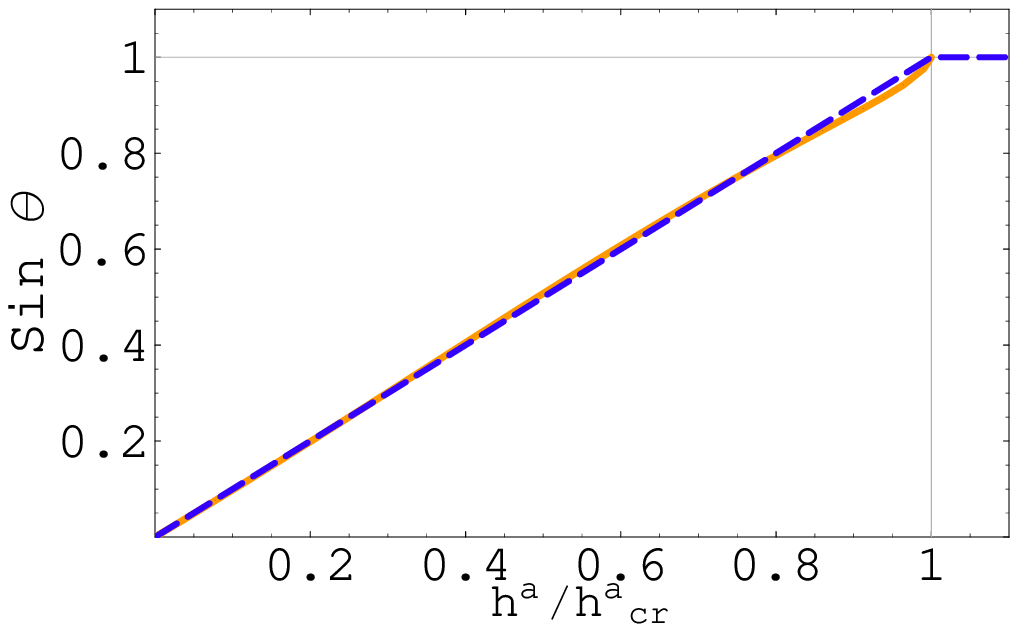}}
\end{picture}}
%\vspace{0.2cm}
%\vspace{0.2cm}
\caption{Ordered moment as a function of transverse magnetic field strength.
 Dashed line: classical behavior; full line: with leading quantum corrections, from
 Eq.~(\ref{reduction}). In the inset, Eq.~\ref{cantingangle} is
 plotted as a function of the transverse field in full line, whereas
 $\sin\theta_0$ is plotted is dashed line.}
\label{Onsite}
\end{figure}

The ordered moment is reduced from its classical value by quantum fluctuations.
At leading order
%===========================================================
\ba
\langle S \rangle \equiv
\langle S^{z}_{\bf R} \rangle &=&S-\frac{1}{N}\sum_{\bf k} \langle \phi^\dagger_{\bf
k} \phi^{\phantom{\dagger}}_{\bf k} \rangle 
\nonumber\\&=& S  - \frac{1}{N}
\sum_{\bf k} \frac{1}{2} \left(\frac{ A_{\bf k}}{\sqrt{A^2_{\bf k}-B^2_{\bf k}}}-1
\right).
\label{reduction}         
\ea
%===========================================================
This is shown  as a function of  transverse magnetic
field in Fig.~\ref{Onsite}. 
%using the renormalized ordering
%wavevector determined from Eq. \ref{RenormalizedQ}.
Our zero-field value of $\langle S \rangle =0.25$ is close to
the result $\langle S \rangle = 0.266+ {\mathcal
O}(S^{-3})$ for the isotropic
triangular antiferromagnet,\cite{Chubukov1}
and to results for the anisotropic lattice without DM interactions,
obtained using the $1/S$ expansion\cite{Trumper,Merino} and series expansions.\cite{Zheng}

The canting angle $\theta$ can be determined
in two different but equivalent ways. Classically, the condition $\theta=\theta_{0}$
ensures both that ${\mathcal H}_{1}=0$ and that $\langle {\mathcal H_0} \rangle$
is at a minimum. The  leading $1/S$ 
correction can be determined similarly. First, normal
ordering of ${\mathcal H}_{3}$, expressed in terms of 
$\gamma^{{\dagger}}_{\bf k}$ and $\gamma^{\phantom{\dagger}}_{\bf k}$,
yields a term linear in boson operators, which should be combined with ${\mathcal H}_{1}$:
the combination vanishes when $\theta$ takes its ground state value.
Second, and alternatively, one can minimize $\langle {\mathcal H}_0+ {\mathcal H}_2\rangle$
with respect to $\theta$. In this way we find
%\be
%\sin \tilde{\theta}= \sin{\theta_{0}} \left[ 1 + \frac{1}{S N}\sum_{\bf k} \frac{1}{2}
%\left( \frac{A_{\bf k}}{\sqrt{A^2_{\bf k}-B^2_{\bf k}}}-1 \right) -
%\frac{1}{S N} \sum_{\bf k} \frac{1/2( J^T_{\bf
%Q+k}+J^T_{\bf Q-k}) -J_{\bf k}}{2 \left[J^T_{\bf 0}-J^T_{\bf Q}\right]}
%\sqrt{\frac{A_{\bf k}-B_{\bf k}}{A_{\bf k}+B_{\bf k}}} \right].
%\ee
\ba
\sin {\theta}= \sin{\theta_{0}} \Bigg[ 1 &+& \frac{1}{2 S N}\sum_{\bf k} 
\left( \frac{A_{\bf k}}{\sqrt{A^2_{\bf k}-B^2_{\bf k}}}-1 \right) 
\nonumber \\&+&
\frac{1}{2 S N} \sum_{\bf k} \frac{B_{\bf k}}{B_{\bf 0}}
\sqrt{\frac{A_{\bf k}-B_{\bf k}}{A_{\bf k}+B_{\bf k}}} \Bigg]\,,
\label{cantingangle}
\ea
where $A_{\bf k}$ and $B_{\bf k}$ should be evaluated at $\theta_0$.
Because quantum fluctuations are suppressed as the critical field is approached,
$\theta \rightarrow \theta_0$ as $h^a \rightarrow
h^a_{cr}$. As seen in the inset of Fig.~\ref{Onsite}, the
quantum corrections to 
$\sin \theta$ are small and Eq.~\ref{cantingangle} is nearly equal to the 
unrenormalized function $\sin\theta_0 ( = h^a/h^a_{cr})$.

%%%%%%%%%%%%%%%%%%%%%%%%%%%%%%%%%%%%%[1]%%%%%%%%%%%%%%%%%%%%%%%%%%%%%%%%%%%%%
%%%%%%%%%%%%%%%%%%%%%%%%%%%%%%%%%%%%%%%%%%%%%%%%%%%%%%%%%%%%%%%%%%%%%%%% 

\begin{figure}[h2t]
\begin{picture}(200,150)
\put(0,0){\includegraphics[width=8cm]{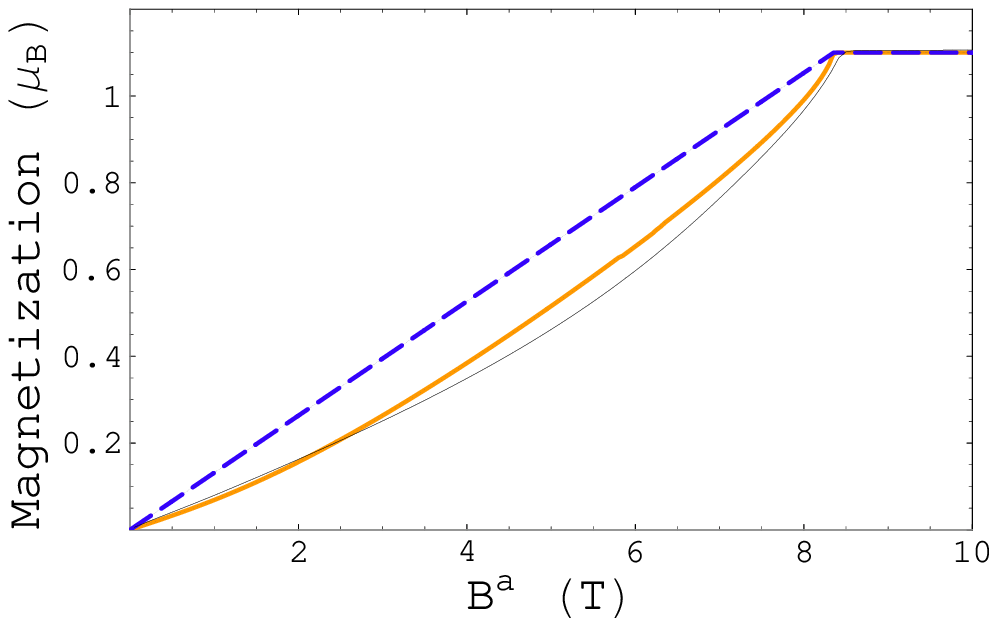}}
\put(33,78){\includegraphics[width=3.5cm]{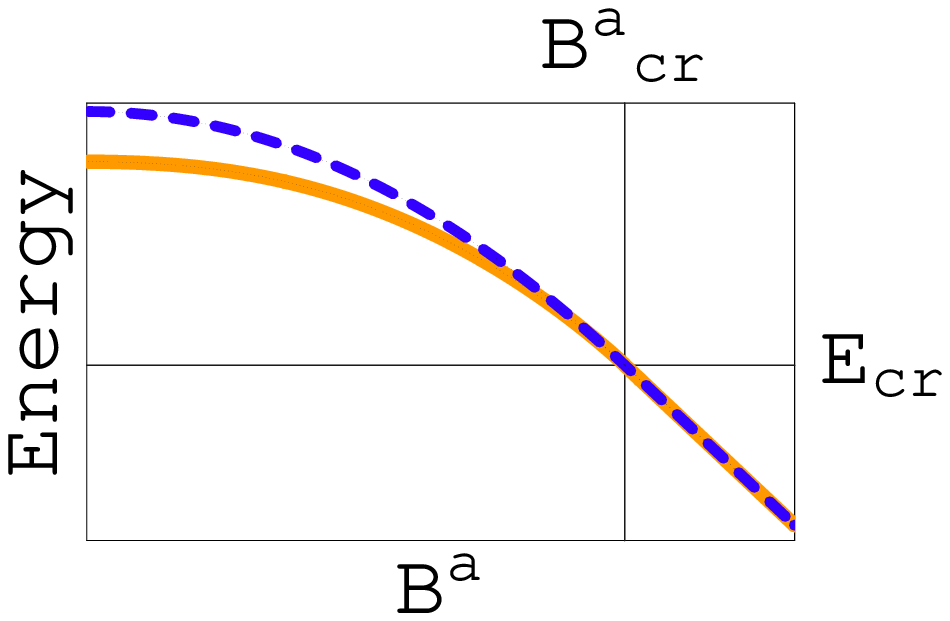}}
\end{picture}
%\vspace{0.2cm}
%\vspace{0.2cm}
\caption{Magnetization as a function of transverse magnetic field strength. 
The dashed line shows the result of classical theory. The thick, full line
includes the $1/S$ correction from Eq.~(\ref{ma}).
The thin, full line gives experimental data,\cite{Private2} measured at $T=60$mk.
Inset: calculated ground state energy as a function of transverse
magnetic field. The dashed and thick lines are
the results of classical theory and the $1/S$ correction,
Eq.~(\ref{RenormalizedQ}), respectively. }
\label{Magnetization}
\end{figure}

Combining results for the ordered moment and the
canting angle, the magnetization is given by
\be
m^a= \frac{g^a \mu_{B}}{N} \sum_{\bf R} \langle S^a_{\bf R} \rangle=
g^a \mu_B \langle S^{z}_{\bf R} \rangle \sin {\theta}\,,
\ee
which yields
%\be
%\langle S^a_{\bf R}\rangle =  \rangle S^{z^\prime}_{\bf R} \sin \theta
%\langle= \rangle S \langle \sin \theta
%which yield,
%===========================================================
%m^a= \frac{ (g^a \mu_B)^2  B^a}{2 J^T_{\bf 0} -2 J^T_{\bf Q}} \left( 1- \frac{1}{N S} \sum_{\bf k} \langle \phi^\dagger_{\bf k} \phi^{\phantom{\dagger}}_{\bf k} \rangle \right).
\be
m^a = \frac{ (g^a \mu_B)^2 B^a}{2 \left[J^T_{\bf 0}-J^T_{\bf Q}
\right]} \left[ 1 + \frac{1}{2 S N} \sum_{\bf k} \frac{B_{\bf k}}{B_{\bf
0}}\sqrt{\frac{A_{\bf k}-B_{\bf k}}{A_{\bf k}+B_{\bf k}}} \right].
\label{ma}
\ee
%m^a(B^a)= \frac{ (g^a \mu_B)^2 B^a}{2 \left[J^T_{\bf 0}-J^T_{\bf Q}
%\right]} \left[ 1 - \frac{1}{S N} \sum_{\bf k} \frac{1/2( J^T_{\bf
%Q+k}+J^T_{\bf Q-k}) -J_{\bf k}}{2 \left[J^T_{\bf 0}-J^T_{\bf Q}\right]}
%\sqrt{\frac{A_{\bf k}-B_{\bf k}}{A_{\bf k}+B_{\bf k}}} \right].
%\label{ma}
%===========================================================
for $h^a < h^a_{cr}$, and $m^a = g^a \mu_B S$ for $h^a > h^a_{cr}$.
The $1/S$ correction on the right-hand side of Eq.~(\ref{ma}) 
has a dependence on the magnetic field through the values of
$A_{\bf k}$ and $B_{\bf k}$ (which again should be evaluated at $\theta_0$).
The departure of the magnetization curve from the
simple linear dependence expected classically is hence a
consequence of zero-point fluctuations.
To understand the sign of this departure, it is useful to recall that
the ground-state energy $E(B^a)$ as a function of field is related to 
the magnetization $m^a(B^a)$ via
\be
E(0) -E(B^a_{cr})= N \int^{B^{a}_{cr}}_{0} m^a(B^a) dB^a\,.
\label{gs}
\ee
Now, since fluctuations reduce $E(0)$ below its classical value but do not contribute
to $E(B^a_{cr})$, the fluctuation contribution to Eq.~(\ref{gs}) is negative.
Supposing this to be true not only for the integral but also for the integrand at all $B^a$,
it is natural to expect the magnetization curve at finite $S$ to lie below the classical one
for all $B^a<B^a_{cr}$. A comparison between our results and
experimental data, presented in Fig.~\ref{Magnetization}, shows a very
good agreement. 

% except above the saturation
%field where the measured $m^a$ is a little larger than expected, and shows a
%small variation with $B^a$.

%Since quantum
%fluctuations are quenched in high magnetic field and the energy of the
%ferromagnetic ground state is merely the classical energy, then on
%general ground, the zero-point motion have to reduce the magnetization
%from its mean field value since the magnetization is the derivative of
%the energy with respect to the field. 
%%%%%%%%%%%%%%%%%%%%%%%%%%%%%%%%%%%%%[1]%%%%%%%%%%%%%%%%%%%%%%%%%%%%%%%%%%%%%
%%%%%%%%%%%%%%%%%%%%%%%%%%%%%%%%%%%%%%%%%%%%%%%%%%%%%%%%%%%%%%%%%%%%%%%% 

The component $S^T$ of the ordered moment in the plane perpendicular to the
applied field can also be evaluated within the $1/S$
expansion. Defining it by
\ba
%\langle {S}^b_{\bf R} \hat{b}+  {S}^c_{\bf R} \hat{c} \rangle
%&=& S^T \left( \cos ({\bf Q}
%\cdot {\bf R}) \hat{b} -(-1)^n \sin({\bf Q} \cdot {\bf R}) \hat{c} \right)\,,
S^T = |\langle {\bf S}_{\bf R} - \hat{\bf a} \langle S_{\bf R}^a \rangle |
\ea
we find $S^T=  \langle S \rangle  \cos {\theta} $. Using
Eqs.~(\ref{reduction}) and (\ref{cantingangle}), we obtain at first order in
$1/S$
\ba
S^T=  S \cos \theta_{0} \Bigg[1&-&
\frac{\sec^2 \theta_{0}}{2 S N} \sum_{\bf k} 
\left( \frac{A_{\bf k}}{\sqrt{A^2_{\bf k}-B^2_{\bf k}}}-1 \right)
\nonumber \\&-&
\frac{\tan^2 \theta_{0}}{2 S N} \sum_{\bf k} \frac{ B_{\bf k}}{B_{\bf 0}}
\sqrt{\frac{A_{\bf k}-B_{\bf k}}{A_{\bf k}+B_{\bf k}}} \Bigg]\,.
\label{ST}
\ea
%===========================================================
%\be
%S^{T}= \sqrt{(S^b)^2+(S^c )^2}= \left( S- \frac{1}{N} \sum_{\bf k}
%\langle \phi^\dagger_{\bf k} \phi^{\phantom{\dagger}}_{\bf k} \rangle
%\right)\sqrt{1 -\left( \frac{h^a}{h^a_{cr}} \right)^2}.
%\ee
%\ba
%S^T&=&  S \sqrt{1- \left( \frac{h^a}{h^a_{cr}}\right) ^2} \left[1-
%\frac{1}{1-\left(\frac{h^a}{h^a_{cr}}\right)^2}
%\frac{1}{S N} \sum_{\bf k} \frac{1}{2}
%\left( \frac{A_{\bf k}}{\sqrt{A^2_{\bf k}-B^2_{\bf k}}}-1 \right)+
%frac{1}{\left( \frac{ h^a_{cr}}{h^a} \right)^2-1} \frac{1}{S N} \sum_{\bf k} \frac{1/2( J^T_{\bf
%Q+k}+J^T_{\bf Q-k}) -J_{\bf k}}{2 \left[J^T_{\bf 0}-J^T_{\bf Q}\right]}
%\sqrt{\frac{A_{\bf k}-B_{\bf k}}{A_{\bf k}+B_{\bf k}}} \right].
%\ea
%===========================================================

\begin{figure}[ht]
\begin{center}
\includegraphics[width=8cm]{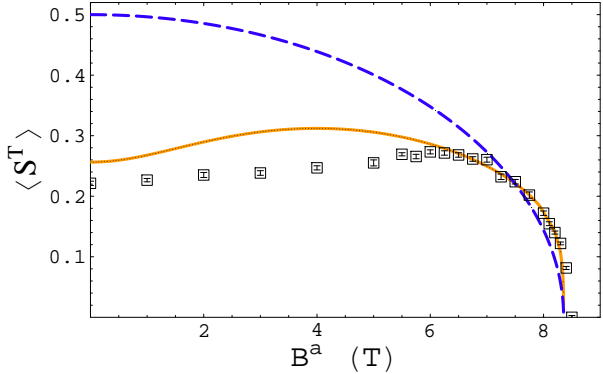} 
%\vspace{0.2cm}
\caption{Component of ordered moment $S^T$ in the plane perpendicular
to the field direction, as a function of transverse magnetic field strength.
Dashed line: classical theory. Full line: result including $1/S$ corrections, from Eq.~(\ref{ST}).
Symbols ($\Box$): experimental data at $T<0.1K$ (Taken from Fig. 3b of Ref.~\onlinecite{Coldea2})}
\label{OrderParameter}
\end{center}
\end{figure}

Results are presented in Fig.~\ref{OrderParameter}.
While classical theory gives $ S^T \propto (1-(B^a/B^a_{cr})^2)^{1/2}$,
fluctuations generate a non-monotonic dependence of $S^T$ on $B^a$ at low fields.
This behavior can be
understood on the basis that polarization of the spins with increasing applied field
has the effect of reducing the
phase space available for quantum fluctuations, and hence increases 
order. Experimental data are also shown in Fig.~\ref{OrderParameter}:
since the absolute scale for $S^T$ has not been determined, we scale
the data to fit theory at high fields. The result of the $1/S$ expansion compares favorably to the
experimental data, which also shows that at low field the
perpendicular ordered moment increases with increasing field .
%%%%%%%%%%%%%%%%%%%%%%%%%%%%%%%%%%%%%%%%%%%%%%%%%%%%%%%%%%%%%%%%%%%%%%%%%%
%%%%%%%%%%%%%%%%%%%%%%%%%%%%%%%%%%%%%%%%%%%%%%%%%%%%%%%%%%%%%%%%%%%%%%%% 

\subsubsection{Longitudinal Field}

In longitudinal field, calculations of fluctuation effects
using the $1/S$ expansion are complicated by the fact that the classical
ground state contains many Fourier harmonics.
At low transverse fields, the classical ground state consists
of a distorted cycloid in which spins lie within the $b$-$c$ plane,
as described in Sec.~\ref{Classical Analysis}. 
In this regime we write 
${\bf S}_{\bf R}= S (0, \cos\phi_{\bf R}, \sin \phi_{\bf R})$ and
consider the leading anharmonic distortion to the cycloid structure,
\cite{Nagamiya0, Zhitomirsky02}
%===========================================================
\be
\phi_{\bf R}= {\bf Q} \cdot {\bf R} + \beta \cos {\bf Q} \cdot {\bf R},
\label{anharmonic}
\ee
%===========================================================
where for concreteness, the field is
taken to be along the $c$ axis. The distortion of the cycloid is
parameterized by $\beta$: its value, determined by minimizing the classical energy, is
%===========================================================
\be
\beta= \frac{h^c}{J^T_{\bf 2 Q}+J^T_{\bf 0}- 2J^T_{\bf Q}}\,.
\ee
%===========================================================

We consider quantum fluctuations about this classical state, using
the Holstein-Primakoff transformation and omitting terms ${\cal O}([h^c]^3)$
and ${\cal O}(S^0)$ to obtain the Hamiltonian
%===========================================================
\be
{\mathcal H}= {\mathcal H}_0+{\mathcal H}_2 ,
\ee
%===========================================================
with
%===========================================================
%\ba 
%{\mathcal H}_0 &=& N S^2 \left( J^{T}_{\bf Q}- \frac{(h^c)^2}{4 \left
%[ J^T_{\bf 2 Q}+ J^{T}_{\bf 0}- 2 J^{T}_{\bf Q} \right]} \right), \\
%{\mathcal H}_2 &=& N S J^{T}_{\bf Q} + S \sum_{\bf k}
%(A^{\prime}_{\bf k} + \beta^2 F_{\bf k}) \left( \phi^\dagger_{\bf k}
%\phi^{\phantom{\dagger}}_{\bf k}+\phi^{\phantom{\dagger}}_{\bf k}
%\phi^{{\dagger}}_{\bf k} \right) + (B^{\prime}_{\bf k} -\beta^2 F_{\bf
%k}) \left
%( \phi^\dagger_{\bf -k} \phi^{{\dagger}}_{\bf
%k}+\phi^{\phantom{\dagger}}_{\bf -k} \phi^{\phantom{\dagger}}_{\bf k}
%\right) \\
%&+& \frac{i S \beta }{8} \sum_{\bf k} G_{\bf k} \left
%( \phi^{\phantom{\dagger}}_{\bf k} \phi^{\phantom{\dagger}}_{-\bf k+Q}-
%\phi^{\phantom{\dagger}}_{\bf -k} \phi^{\phantom{\dagger}}_{\bf k-Q} - h.c.  \right) 
%+ \left[ G_{\bf
%k}+4(J^T_{\bf 2Q}-J^T_{\bf Q}) \right]\left( \phi^{\dagger}_{\bf k} \phi^{\phantom{\dagger}}_{\bf k-Q}-
%\phi^{\dagger}_{\bf -k} \phi^{\phantom{\dagger}}_{\bf k-Q} - h.c.
%\right).
%\ea
\ba 
{\mathcal H}_0 &=& N S^2 \left( J^{T}_{\bf Q}- \frac{(h^c)^2}{4 \left
[ J^T_{\bf 2 Q}+ J^{T}_{\bf 0}- 2 J^{T}_{\bf Q} \right]} \right) 
%\nonumber\\
\ea
and
\begin{widetext}
\ba
{\mathcal H}_2 &=& N S J^{T}_{\bf Q} + S \sum_{\bf k}\bigg[
A^{\prime}_{\bf k}  \left( \phi^\dagger_{\bf k}
\phi^{\phantom{\dagger}}_{\bf k}+\phi^{\phantom{\dagger}}_{\bf k}
\phi^{{\dagger}}_{\bf k} \right)
%\nonumber \\ &+& 
+B^{\prime}_{\bf k} \left
( \phi^\dagger_{\bf -k} \phi^{{\dagger}}_{\bf
k}+\phi^{\phantom{\dagger}}_{\bf -k} \phi^{\phantom{\dagger}}_{\bf k}
\right)\bigg] \nonumber \\
&+& \frac{i S \beta }{2} \sum_{\bf k} \bigg[L_{\bf k} \left
( \phi^{\phantom{\dagger}}_{\bf k} \phi^{\phantom{\dagger}}_{-\bf k+Q}-
\phi^{\phantom{\dagger}}_{\bf -k} \phi^{\phantom{\dagger}}_{\bf k-Q} - h.c.  \right) 
%\nonumber \\
%+&& \!\!\!\!\!\!\!
+\left[ L_{\bf
k}+J^T_{\bf 2Q}-J^T_{\bf Q} \right]\left( \phi^{\dagger}_{\bf k} \phi^{\phantom{\dagger}}_{\bf k-Q}-
\phi^{\dagger}_{\bf -k} \phi^{\phantom{\dagger}}_{\bf k-Q} - h.c.
\right)\bigg]\,.
\ea

%===========================================================
Here
%===========================================================
\ba
 A^{\prime}_{\bf k}&=& \frac{1}{4} \left[ 2 J_{\bf k}+ J^{T}_{\bf Q+k}+ J^{T}_{\bf
 Q-k}\right] -J^{T}_{\bf Q}   +\beta^2 \left( \frac{J^T_{\bf
 2Q+k}+J^{T}_{\bf 2Q - k}+J^T_{\bf k}+J^{T}_{\bf - k}}{16}-
\frac{J^T_{\bf Q+k}+J^{T}_{\bf Q - k}}{8} \right)\,, \\
 B^{\prime}_{\bf k}&=& \frac{1}{4} \left[ 2 J_{\bf k}-J^{T}_{\bf Q + k}- J^{T}_{\bf
 Q - k}\right] - \beta^2 \left( \frac{J^T_{\bf 2Q+k}+J^{T}_{\bf 2Q -
 k}+J^T_{\bf k}+J^{T}_{\bf - k}}{16}-
\frac{J^T_{\bf Q+k}+J^{T}_{\bf Q - k}}{8} \right)\,, \\
L_{\bf k}&=& \frac{1}{8} \left( -J^T_{\bf 3Q -k}+J^T_{\bf -3Q+k}-2J^T_{\bf
-2Q +k} -J^T_{\bf Q+k} -J^T_{\bf -Q-k} + 2 J^T_{\bf -k} +2 J^T_{\bf -Q+k} \right)\,.
\ea

%\ba
% A^{\prime}_{\bf k}&=& \frac{1}{4} \left[ 2 J_{\bf k}+ J^{T}_{\bf Q+k}+ J^{T}_{\bf
% Q-k}\right] -J^{T}_{\bf Q}  , \\
% B^{\prime}_{\bf k}&=& \frac{1}{4} \left[ 2 J_{\bf k}-J^{T}_{\bf Q + k}- J^{T}_{\bf
% Q - k}\right]  , \\
%F_{\bf k}&=& \left[ \frac{J^T_{\bf
% 2Q+k}+J^{T}_{\bf 2Q - k}+J^T_{\bf k}+J^{T}_{\bf - k}}{16}-
%\frac{J^T_{\bf Q+k}+J^{T}_{\bf Q - k}}{8} \right], \\
%G_{\bf k}&=& \frac{1}{2} \left[ -J^T_{\bf 3Q -k}+J^T_{\bf -3Q+k}-2J^T_{\bf
%-2Q +k} -J^T_{\bf Q+k} -J^T_{\bf -Q-k} + 2 J^T_{\bf -k} +2 J^T_{\bf
%-Q+k} \right].
%\ea
\end{widetext}

The higher Fourier harmonics in the classical ground state
scatter spin fluctuations with a momentum transfer which is a multiple of ${\bf Q}$.
The presence of these scattering terms, proportional to $\beta$, in the quadratic 
spinwave Hamiltonian means that the dispersion relation is determined
 by an infinite set of coupled equations.\cite{Ziman}
Since our calculation is anyway restricted to small $h^c \propto \beta$,
we treat these coupled equations to ${\cal O}(\beta^2)$
in a calculation of the ground state energy.
More specifically, it is convenient  first to perform a Bogoliubov transformation to 
diagonalize the momentum conserving terms, and then
to evaluate the contribution from the terms scattering by $\pm {\bf Q}$
using perturbation theory. We find
%===========================================================
\ba
E&=&N S (S+1) J^{T}_{\bf Q} - \frac{N S^2 (h^c)^2}{4 \left[ J^T_{\bf 2
Q}+ J^{T}_{\bf 0}- 2 J^{T}_{\bf Q} \right]} 
\nonumber \\&+& S \sum_{\bf k}
\Omega^{\prime}_{\bf k}-  S \beta^2 \sum_{\bf k }  \frac{ 2 | I_{\bf k
 }|^2}{\Omega^{\prime}_{\bf k} +\Omega^{\prime}_{\bf Q - k}},  
\label{LowFieldE}
\ea
with $\Omega^{\prime}_{\bf k}= \sqrt{(A^{\prime}_{\bf
k})^2-(B^{\prime}_{\bf k})^2
}$ and 
\ba
I_{\bf k}= &-&\frac{i}{2} \big[ L_{\bf k} (u^{\prime}_{\bf k} -v^{\prime}_{\bf k})
(u^{\prime}_{\bf Q-k} -v^{\prime}_{\bf Q-k}) 
\nonumber \\&-&( J^T_{\bf 2 Q}-J^T_{\bf Q}) (u^{\prime}_{\bf k} v^{\prime}_{\bf Q-k}+v^{\prime}_{\bf k} u^{\prime}_{\bf Q-k}) \big],
\ea
%===========================================================
where $u^{\prime}_{\bf k}$ and $v^{\prime}_{\bf k}$ are given by
Eq.~(\ref{Bogoliubov2}) after substituting $A^{\prime}_{\bf k}$ and $B^{\prime}_{\bf k}$
for $A_{\bf k}$ and $B_{\bf k}$.

\begin{figure}[ht]
\begin{center}
\includegraphics[width=8cm]{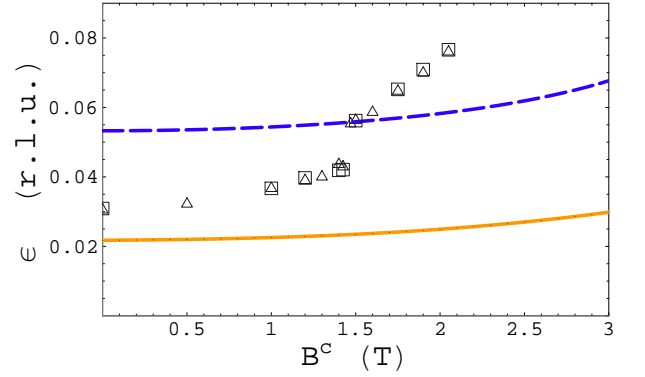} 
%\vspace{0.2cm}
\caption{
  The incommensuration $\epsilon$ as a function of longitudinal field strength.
  Solid line: result from $1/S$ expansion, from
  Eq.~(\ref{LowFieldE}). Dashed line: result from the classical result.
  Symbols are the experimental results taken from
  Ref.~\onlinecite{Coldea1}, Fig. 1e: $\triangle$ from magnetic peaks at
  ${\bf Q} =(2,1/2+\epsilon,0)$ and $\Box$ from peaks at ${\bf
  Q}=(1,1/2+\epsilon,0)$. Experimentally, a distorted cycloid occurs for
  $B^c<1.4$T and a second incommensurate phase occurs for 
  fields in the range $ 1.4 {\rm T}<B^c<2.1 {\rm T}$. }
\label{LongIncLF}
\end{center}
\end{figure}

The ordering wavevector and its dependence on field can be calculated by
minimizing Eq.~(\ref{LowFieldE}) with respect to ${\bf Q}$. 
It is interesting to note that, in contrast to the case for a transverse field, 
the ordering wavevector in a longitudinal field is dependent on field strength 
even at the classical level. 
It increases with field and this trend is  
reinforced by the quantum fluctuations. Results are shown in
Fig.~\ref{LongIncLF}, together with experimental data.
The observed increase of ${\bf Q}$ with field is much faster than
the calculated one; 
the origin of this discrepancy is not understood.

%%%%%%%%%%%%%%%%%%%%%%%%%%%%%%%%%%%%%%%%%%%%%%%%%%%%%%%%%%%%%%%%%%%%%%%%%%
%%%%%%%%%%%%%%%%%%%%%%%%%%%%%%%%%%%%%%%%%%%%%%%%%%%%%%%%%%%%%%%%%%%%%%%% 

%For completeness, we give the magnetization at $T=0$, calculated by
%determining of the ground state energy with respect to the magnetic
%field,
%%===========================================================
%\be
%m^c= \frac{ (g^c \mu_B)^2 B^c}{2 J^T_{\bf 2 Q} + 2J^T_{\bf 0} -4
%J^T_{\bf Q}} \left[ 1 - \frac{1}{2 S N} \sum_{\bf k} \frac{1/2(
%J^T_{\bf 2Q+k}+J^T_{\bf 2Q-k}+J^T_{\bf k}+J^T_{\bf -k}) -J^T_{\bf
%Q+k}-J^T_{\bf Q-k}}{J^T_{\bf 2 Q}+J^T_{\bf 0 }-2 J^T_{\bf Q}}
%\sqrt{\frac{A^{\prime}_{\bf k}+B^{\prime}_{\bf k}}{A^{\prime}_{\bf
%k}-B^{\prime}_{\bf k}}} \right].
%\ee
%%===========================================================
%We stress that this result holds in low longitudinal field regime where the
%anharmonicities induced by the fields are small. 

\begin{figure}[ht]
\begin{center}
\includegraphics[width=8cm]{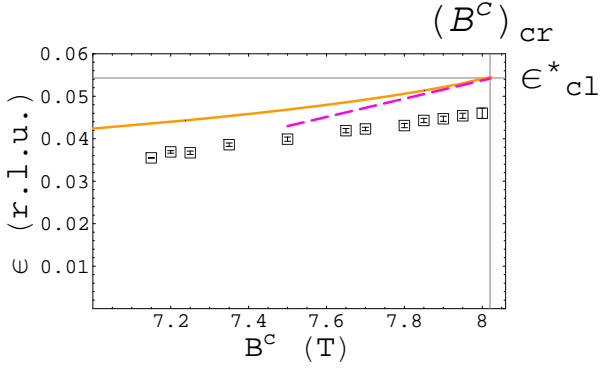} 
%\vspace{0.2cm}
\caption{The incommensuration $\epsilon$ as a function of longitudinal magnetic
  field strength $B^c$. Solid line: result from $1/S$ expansion. Dashed line: result from calculation
  for dilute Bose gas of reversed spins. Symbols ($\Box$): experimental data\cite{Coldea4}
  (no incommensurate ordering is observed for 2.1 T$<B^c<$7.1 T).}
\label{LongitudinalIncommensuration}
\end{center}
\end{figure}

Now we turn to the case of stronger longitudinal fields. 
In the field range $0.35 < h^c/h^c_{cr} < 1$, the classical treatment
described in Sec.~\ref{Classical Analysis} gives the tilted cone of
Eq.~(\ref{tilted-cone}) as the ground state.
The tilting angle $\eta$ of the cone axis from the field direction
is given approximately by $\tan \eta \approx 
({ D }/{h^c_{cr}}) (1-\left({h^c_{cr}}/{h^c}\right)^2)$ 
and is less than one degree for $h^c > 0.9 h^c_{cr}$.
Experimentally, an incommensurately ordered state has recently been observed\cite{Coldea4} for
$7.1 {\rm T} < B^c < B^c_{cr}$.
With this in mind, we approximate the classical ground state in this field range
by setting $\eta = 0$ in Eq.~(\ref{tilted-cone}) and use the $1/S$
expansion to study the effects of quantum fluctuations.
Following a procedure similar to the one described for a
transverse field, we have calculated the ordering wavevector
as a function of field.
% Experimentally, an incommensurate ordered
%phase has been reported recently\cite{Coldea4} for $7.1T < B^c < B^c_{cr}$. 
While the observed phase has not so far been fully characterized, its ordering wavevector 
has been measured as a function of field strength.
From our classical calculation, we expect the state to be
a tilted cone. Our results for the ordering wavevector are
compared with experimental data in Fig.~\ref{LongitudinalIncommensuration}.
Calculated and observed values of the ordering wavevector vary in
the same way with field, but there is an offset between the
two which remains a puzzle. In the following subsection (\ref{Dilute Bose Gas}),
we obtain results which are essentially exact close to $B^c_{cr}$.
Since the discrepancy remains, we conclude that the value of the ordering wavevector is
influenced by interactions not
included in the model Hamiltonian of Eq.~(\ref{Hamiltonian00}).
%%%%%%%%%%%%%%%%%%%%%%%%%%%%%%%%%%%%%%%%%%%%%%%%%%%%%%%%%%%%%%%%%%%%%%%%%%
%%%%%%%%%%%%%%%%%%%%%%%%%%%%%%%%%%%%%%%%%%%%%%%%%%%%%%%%%%%%%%%%%%%%%%%% 

\subsection{Dilute Bose Gas}
\label{Dilute Bose Gas}

An alternative to the $1/S$ expansion can be motivated by noting that 
fully polarized states are exact eigenstates of the Hamiltonian. 
The absence of quantum fluctuations 
suggests a systematic expansion in powers of the density of reversed spins,\cite{Gluzman,Nikuni95}
equivalent to an expansion in powers of $(1-h/h_{cr})$. In this approach
reversed spins constitute a dilute gas of bosons with hard-core repulsion.

\subsubsection{Transverse Field}

We introduce boson 
creation and annihilation operators, $\phi^{{\dagger}}_{\bf R}$ and $\phi^{\phantom{\dagger}}_{\bf R}$
to represent spins, setting, for a transverse field,
\ba
S^{a}_{\bf R}&=& \frac{1}{2} - \phi^{\dagger}_{\bf R}
\phi^{\phantom{\dagger}}_{\bf R}, 
%\qquad 
\nonumber \\
S^{+}_{\bf R}&=& S^{b}_{\bf R}+i S^{c}_{\bf R}=  \phi^{\phantom{\dagger}}_{\bf R}, 
%\qquad
\nonumber\\
S^{-}_{\bf R}&=& S^{b}_{\bf R}-i S^{c}_{\bf R}= \phi^{{\dagger}}_{\bf R}
\label{Hard-coreRepresentation}
\ea
%===========================================================
%\be
%S^{a}_{\bf R}= \frac{1}{2} - \phi^{\dagger}_{\bf R}
%\phi^{\phantom{\dagger}}_{\bf R}, \qquad S^{+}_{\bf R}= S^{b}_{\bf
%R}+i e^{i \pi R^a} S^{c}_{\bf R}= \phi^{\phantom{\dagger}}_{\bf R},
%\qquad S^{-}_{\bf R}= S^{b}_{\bf R}-i e^{i \pi R^a} S^{c}_{\bf R}=
%\phi^{{\dagger}}_{\bf R},
%\ee
%===========================================================
with the constraint that the particle number $n_{\bf
R}=\phi^{\dagger}_{\bf R} \phi^{\phantom{\dagger}}_{\bf R}$ 
takes only the values 0 and 1. This is imposed
by introducing an on-site
interaction $U$ and taking the limit $U
\rightarrow \infty$. %We have
%included an exponential factors in the transformation to unwind the
%alternating $DM$ vector on even and odd layers in the hardcore
%representation.

The Hamiltonian for a single layer (with, for definiteness, the layer index $n$ chosen to be even) 
is
%===========================================================
\ba
{\mathcal H}=& N &\left[\frac{ J_{\bf 0}-h^a}{4} \right] + \sum_{\bf k}
(\epsilon^T_{\bf k} -\mu)\phi^\dagger_{\bf k}
\phi^{\phantom{\dagger}}_{\bf k} 
\nonumber\\ &+& \frac{1}{2 N} \sum_{\bf k,
k^{\prime}, q} V_{\bf q}({\bf k},{\bf k^{\prime}}) \phi^\dagger_{\bf
k+q} \phi^\dagger_{\bf k^{\prime}-q} \phi^{\phantom{\dagger}}_{\bf
k^{\prime}}\phi^{\phantom{\dagger}}_{\bf k},
\ea
%===========================================================
where
%===========================================================
\ba
\epsilon^T_{\bf k} &=&  J^T_{\bf k} - J^T_{\bf Q} , \\
\mu &=& \frac{h^a_{cr} -h^a}{2}, \\
V_{\bf q}( {\bf k, k^{\prime}}) &=& 2 J_{\bf q}+ 2 U .
\ea
%===========================================================

Standard techniques\cite{Abrikosov} developed for the interacting Bose gas can be applied to treat this Hamiltonian.
For $h^a>h^a_{cr}$ the spin system is fully polarized. Equivalently,
for $\mu<0$ the ground state is the boson vacuum. 
Conservation of boson number follows from
U(1) symmetry of the spin Hamiltonian in a transverse field.
Magnetic order of in-plane spin components below the critical field
translates, using Eq.~(\ref{Hard-coreRepresentation}), to formation
of a Bose condensate for $\mu>0$. We introduce the order parameter $\psi_{\bf q}$ and
shift the boson annihilation operator by a constant
%===========================================================
\be
\phi_{ \bf Q} \rightarrow  \sqrt{N} \psi_{ \bf Q} + \phi_{ \bf Q}\,,
\ee
%===========================================================
where the ordering wavevector ${\bf Q}$ need not take the
classical value ${\bf Q}_{cl}$. Minimization of the ground-state energy is equivalent to requiring
$\langle \phi_{ \bf Q} \rangle=0$. 
Working in the low density limit,
the scattering amplitude between bosons
is given by an effective interaction potential $\Gamma_{\bf q} ({\bf
k, k^{\prime}})$ which results from summing  ladder
diagrams. It satisfies the integral equation
%===========================================================
\be
\Gamma_{\bf q} ({\bf k,k^{\prime}})\!\!=\!\! V_{\bf q} - \frac{1}{N} \sum_{\bf q^{\prime}} \frac{ V_{\bf q- q^{\prime}}}{\epsilon^T_{k+q^{\prime}}+\epsilon^T_{k^{\prime}-q^{\prime}}-\epsilon^T_{k}-\epsilon^T_{k^{\prime}}} \Gamma_{\bf q^{\prime}} ({\bf k, k^{\prime}}).
\label{Gamma}
\ee
%===========================================================

The ground state energy, including the leading interaction effects at low density, is
%===========================================================
\ba
E^{(2)}\!\!\!=\!\!N \bigg[ \frac{J^T_{\bf 0}-h^a}{4}\!\!\!
&+&\!\!\!\left(\epsilon^T_{\bf Q} -\mu \right) |\psi_{\bf Q}|^2 
\!\!+ \frac{1}{2}
\Gamma_{\bf 0}\left({\bf Q,Q} \right) |\psi_{\bf Q}|^4 \bigg] 
\nonumber \ \\&+&
\frac{1}{2} \sum_{\bf k} \left( E_{\bf k}  - F_{\bf k} \right)
\label{E2}
\ea
%===========================================================
where $E_{\bf k}= \sqrt{F^2_{\bf k}-G^2_{\bf k}}+N_{\bf k}$, and
%===========================================================
\ba
F_{\bf k} &=& \frac{\epsilon^T_{\bf Q+k}+\epsilon^T_{\bf Q-k}}{2}
-\mu +\frac{|\psi_{ \bf Q}|^2}{2} \bigg[
\Gamma_{\bf k} ({\bf Q, Q+k})
\nonumber\\
+&&\!\!\!\!\!\Gamma_{\bf -k} ({\bf Q+k, Q})+\Gamma_{\bf 0} ({\bf Q, Q+k})+\Gamma_{\bf 0} ({\bf Q+k, Q}) \bigg], \nonumber \\
G_{\bf k} &=& |\psi_{ \bf Q}|^2 \Gamma_{\bf k} ({\bf Q, Q})\,,
\nonumber \\ N_{\bf
k} &=& \frac{\epsilon^T_{\bf Q+k}-\epsilon^T_{\bf Q-k}}{2}\,.
\ea
%===========================================================
%In Eq. \ref{E2},  represents the energy of the
%particles in the condensate whereas the last term is  the energy
%of the particles out of the condensate (spin wave contribution) in the
%low density limit. 
The condition $\langle \phi_{ \bf Q} \rangle =0$ yields an expression for the order parameter
%===========================================================
\be
|\psi_{ \bf Q}|^2 = \frac{\mu-\epsilon^T_{ \bf Q}}{\Gamma_{\bf 0}
 \left( {\bf Q,Q} \right)}\,. 
\ee
%===========================================================
Substituting this into Eq.~(\ref{E2}), the ground-state energy is
%===========================================================
\be
E^{(2)}= N \left[ \frac{J^T_{\bf 0}}{4}- \frac{h^a}{4} -
\frac{\left(\mu -\epsilon^{T}_{\bf Q} \right)^2}{2 \Gamma_{\bf 0} ({\bf
Q, Q})} \right] + \frac{1}{2} \sum_{\bf k} \left( E_{\bf k} - F_{\bf
k} \right)\,.\nonumber
\ee
%===========================================================

As in the $1/S$ calculation, the
ordering wavevector can be determined as a function of field
by minimizing $E^{(2)}$ with respect to ${\bf Q}$. 
Our focus here, however is on exact results
close to the critical field. At $h^a=h^a_{cr}$ we find ${\bf Q} = {\bf Q}_{cl}$.
In addition, we obtain
%${\bf Q}(\mu)={\bf Q}_{cl} + \hat{e}_{i} \alpha_{i} \mu$
%===========================================================
%\be
%\alpha_{i}= 
%\frac{ \frac{1}{ \Gamma^2_{\bf 0} ({\bf Q},{\bf Q})} \frac{\partial
%\Gamma_{0}({\bf Q},{\bf Q})}{ \partial Q_{i}} - 
%\frac{\partial}{\partial Q_{i}} 
%\sum_{\bf k} 
% \frac{1}{\epsilon_{\bf Q+k}+\epsilon_{\bf Q-k}}
%\left(       \frac{\Gamma_{\bf k}({\bf Q,Q})}{\Gamma_{\bf 0} ({\bf Q,Q})} 
%\right)^2}
%{ \left[\frac{1}{\Gamma_{0} ({\bf Q,Q})} +
%\sum_{\bf k} 
% \frac{1}{\epsilon_{\bf Q+k}+\epsilon_{\bf Q-k}}
%\left(\frac{\Gamma_{\bf k}({\bf Q,Q})}{\Gamma_{\bf 0} ({\bf Q,Q})}
% \right)^2 \right]
%\frac{\partial^2 \epsilon_{\bf Q}}{\partial {Q_{i}}^2}},
%\label{alpha}
%\ee
%===========================================================
%===========================================================
\be
\left.\frac{\partial {\bf Q}}{\partial h^a}\right|_{h^a=h^a_{cr}}
%\alpha_{i}
%= - \frac{1}{2} \left( \frac{\partial^3 E^{(2)}}{\partial Q^2_i \partial \mu} \right)^{-1} 
%\frac{\partial^3 E^{(2)}}{\partial Q_i \partial \mu^2},
\!\!\!\!\!\!\!= \frac{1}{4} \hat{\bf k}\left. \left( \frac{\partial^3 E^{(2)}}{\partial Q^2_k \partial \mu} \right)^{-1} 
\frac{\partial^3 E^{(2)}}{\partial Q_k \partial \mu^2}\right|_{h^a=h^a_{cr},{\bf Q}={\bf Q}_{cl}} \nonumber
\ee
and hence
\begin{widetext}
\be
\left.\frac{\partial {\bf Q}}{\partial h^a}\right|_{h^a=h^a_{cr}}
%\alpha_{i}= \frac{1}{2}
%\left(\frac{\partial^2 \epsilon^{T}_{\bf Q}}{\partial {Q_{i}}^2} \right)^{-1}
%\frac{\partial
%\ln  \left[\frac{1}{\Gamma_{0} ({\bf Q,Q})} +
%\frac{1}{N}\sum_{\bf k} 
% \frac{1}{\epsilon^T_{\bf Q+k}+\epsilon^T_{\bf Q-k}}
%\left(\frac{\Gamma_{\bf k}({\bf Q,Q})}{\Gamma_{\bf 0} ({\bf Q,Q})}
% \right)^2 \right] }{\partial Q_{i}},
= \frac{1}{4}\hat{\bf k}\left.
\left(\frac{\partial^2 \epsilon^{T}_{\bf Q}}{\partial {Q_{k}}^2} \right)^{-1}
\frac{\partial}{\partial Q_k}
\ln  \left[\frac{1}{\Gamma_{0} ({\bf Q,Q})} +
\frac{1}{N}\sum_{\bf k} 
\frac{1}{\epsilon^T_{\bf Q+k}+\epsilon^T_{\bf Q-k}}
\left(\frac{\Gamma_{\bf k}({\bf Q,Q})}{\Gamma_{\bf 0} ({\bf Q,Q})}
\right)^2 \right] \right|_{{\bf Q}={\bf Q}_{cl}}\,.
\label{alpha}
\ee
\end{widetext}
%===========================================================
%where the derivatives are evaluated at the wavevector ${\bf Q}_{cl}$ and $\mu=0$. 
A potential difficulty arises at this point because interactions are marginally
irrelevant at the critical point of the two-dimensional Bose gas:\cite{Fisher01}
in consequence, $\Gamma_{\bf 0} ({\bf Q}_{cl},{\bf Q}_{cl})$
vanishes for an isolated layer. It is therefore essential to include
interlayer exchange $J^{\prime \prime}$ in the calculation of the vertex.
Evaluating Eq.~(\ref{alpha}) numerically, we find $h^a_{cr} \partial {\bf Q}/\partial h^a = (0,0.0911,0)$.
This result is displayed in Fig.~\ref{TransverseIncommensuration}: it is
similar to that given by the $1/S$ expansion, indicating that the linear
spinwave theory captures the effects of quantum fluctuations quite accurately in this system.
Both approaches are in good agreement with experiment, especially close to the critical field.

\subsubsection{Longitudinal Field}

A similar procedure can be followed for the system in longitudinal field (chosen
along the $c$ axis, without loss of generality for the Hamiltonian of Eq.~(\ref{Hamiltonian00})). 
With this field orientation,
the expressions for spin operators in terms of Bose operators are
%===========================================================
%\be
%S^{c}_{\bf R}= \frac{1}{2} - \phi^{\dagger}_{\bf R}
%\phi^{\phantom{\dagger}}_{\bf R}, \qquad S^{+}_{\bf R}= S^{a}_{\bf
%R}+i e^{i \pi R^a} S^{b}_{\bf R}= \phi^{\phantom{\dagger}}_{\bf R},
%\qquad S^{-}_{\bf R}= S^{a}_{\bf R}-i e^{i \pi R^a} S^{b}_{\bf R}=
%\phi^{{\dagger}}_{\bf R}\,.
%\label{Hard-coreRepresentation02}
%\ee
%===========================================================
%===========================================================
\ba
S^{c}_{\bf R}&=& \frac{1}{2} - \phi^{\dagger}_{\bf R}
\phi^{\phantom{\dagger}}_{\bf R}, \nonumber\\
%\qquad 
S^{+}_{\bf R}&=& S^{a}_{\bf
R}+i S^{b}_{\bf R}= \phi^{\phantom{\dagger}}_{\bf R},\nonumber\\
%\qquad 
S^{-}_{\bf R}&=& S^{a}_{\bf R}-i S^{b}_{\bf R}=
\phi^{{\dagger}}_{\bf R}.
\label{Hard-coreRepresentation02}
\ea
%===========================================================
The Hamiltonian for a single layer (again taking the layer index $n$ to be even) is
%===========================================================
\ba
{\mathcal H}&=& N \left[ \frac{J_{\bf 0} - h^c}{4} \right] +\sum_{\bf k}
(\epsilon^L_{\bf k} -\mu)\phi^\dagger_{\bf k}
\phi^{\phantom{\dagger}}_{\bf k} 
\nonumber\\&+& \frac{1}{2 \sqrt{N}} \sum_{\bf k,
k^\prime} \left[ D_{\bf k}+D_{\bf k^\prime} \right] \left
( \phi^{\dagger}_{\bf k+ k^{\prime}} \phi^{\phantom{\dagger}}_{\bf
k^{\prime}}\phi^{\phantom{\dagger}}_{\bf k} +h.c \right) \nonumber\\&+& \frac{1}{2
N} \sum_{\bf k, k^{\prime}, q} V_{\bf q}({\bf k},{\bf k^{\prime}})
\phi^\dagger_{\bf k+q} \phi^\dagger_{\bf k^{\prime}-q}
\phi^{\phantom{\dagger}}_{\bf k^{\prime}}\phi^{\phantom{\dagger}}_{\bf
k},
\ea
%===========================================================
where
%===========================================================
\ba
%\epsilon^L_{\bf k} &=&  J_{\bf k}- J_{\bf Q} , \\
% D_{\bf k}&=& -2D \sin (\pi k) \cos (\pi l), \\
%\mu &=& J_{\bf 0} - J_{\bf Q}  -\frac{h^c}{2}, \\
\epsilon^L_{\bf k} -\mu  &=& J_{\bf k} -J_{\bf 0} +h^c/2 
% V_{\bf q}( {\bf k, k^{\prime}}) &=& 2 J_{\bf q}+ 2 U ,
\ea
%===========================================================

The presence of a term cubic in boson operators considerably complicates the
analysis, since with it, particle number is not conserved.
Its appearance reflects the fact that a longitudinal field breaks U(1)
symmetry as discussed in Sec. \ref{Spin Hamiltonian}. In general, the
particle number (or longitudinal magnetization) is not conserved
(except for the boson vacuum  -- the ferromagnetic state -- which is an exact eigenstate of the Hamiltonian).
The remaining, ${\mathbb Z}_2$ symmetry is invariant under
the canonical transformation
$\phi^{\dagger}_{\bf k} \rightarrow -\phi^{\dagger}_{\bf -k}$. 

We note in passing that the cubic term does not result in a first
order transition from the fully polarized state as $\mu$ is varied, because momentum conservation precludes contributions involving only the
ordering fields $\phi_{\bf Q}$ and $\phi_{\bf -Q}$ in a
Landau-Ginzburg description. An ordered state is therefore brought
about by the closing of the single-particle excitation gap, yielding
a second order phase transition. The universality class associated with this
quantum phase transition must take into account the
 extra ${\mathbb Z}_2$ symmetry of the Hamiltonian. The low energy action
is described by a ${\mathbb Z}_2 \oplus U(1) $ symmetry  model. This multicritical transition found in
longitudinal field is to be contrasted with the ordinary XY quantum
phase transition found in transverse field. \cite{Pelisetto} 

While for a transverse field the ordering wavevector 
can be found simply from the quadratic part of the boson Hamiltonian, this is not so 
for a longitudinal field. In that case, because particle number is not conserved, 
the quasiparticle spectrum is renormalized by quantum fluctuations, 
even at the critical point. It is interesting to check whether
a renormalization of this kind may be responsible for the
discrepancy between theory and experiment shown in Fig.~\ref{LongitudinalIncommensuration}.
The critical field and the
ordering wavevector are determined from the
values of $h^c$ and ${\bf Q}$ for which the one-particle Green function
has a pole at zero energy, by solving 
%===========================================================
\be
G\left({\bf Q}, E_{\bf Q}=0 \right)^{-1}=0.
\ee
%===========================================================
In absence of DM interactions, the
one-particle Green function at and above the critical field is given 
exactly at zero temperature by
the expression for a non-interacting system, $G^0 \left( {\bf k}, i \omega
\right)=({i \omega -\epsilon^L_{\bf k} + \mu})^{-1}$. Since $D\ll J$,
we evaluate the leading
contribution to the self-energy,
%===========================================================
\be
\Sigma \left( {\bf k}, i \omega \right)= \frac{1}{2 N} \sum_{\bf q} \frac{ \left[ D_{\frac{\bf k+q}{2}}+ D_{\frac{\bf k-q}{2}}\right]^2}{i  \omega - \epsilon^{L}_{\frac{\bf k+q}{2}}-\epsilon^{L}_{\frac{\bf k-q}{2}}+ 2\mu}
 + {\mathcal O}\left( D^4 \right).
\ee
%===========================================================
yielding a renormalized quasiparticle spectrum $ \omega_{\bf k} \simeq
\epsilon^L_{\bf k}-\mu+ \Sigma({\bf k}, \epsilon^L_{\bf k}-\mu)$ in the
symmetric phase. The ordering wavevector can be found by solving Dyson's equation at
the critical field,
%===========================================================
\be
G^0\left({\bf Q},0 \right)^{-1}- \Sigma \left( {\bf Q}, 0 \right)=0,
\ee
%===========================================================
which gives ${\bf Q}= {\bf Q}^{\star}_{cl}+ (0, 0.00025,0)$. 
This minute quantum correction at the
critical field is nearly two orders of magnitude too small to explain
the discrepancy between ${\bf Q}^{\star}_{cl}$ and the experimental
ordering wavevector illustrated in Fig.~\ref{LongitudinalIncommensuration}. 
We conclude that there are further anisotropic interactions
present in the system but not captured by the Hamiltonian of Eq.~(\ref{Hamiltonian00}).
Additional evidence for this is provided by the fact that the experimental 
phase diagram in a longitudinal field depends on field orientation
{\em within} the $b$-$c$ plane.\cite{Coldea4}

It is interesting to note that 
at the critical point in a longitudinal field,
in contrast to behavior for a transverse field,
order is possible at two wavevectors, $\pm {\bf Q}$.
Cone states
break spontaneously the Ising symmetry, with condensation either at
${\bf Q}$ or at $-{\bf Q}$.
An alternative possibility is the {\em simultaneous} condensation of magnons at
both wavevectors, forming a fan phase.
Competition
between the fan and cone phases is determined by the
interaction between magnons. A straightforward calculation shows that
the cone phase is favored if
%===========================================================
\be
\Gamma_{\bf 0}({\bf Q, Q}) < \Gamma_{\bf 0}({\bf Q, -Q}) + \Gamma_{\bf 2 Q}({\bf Q, -Q}),
\ee
%===========================================================
while the fan phase is preferred if the inequality is reversed. 
Evaluating the vertices numerically, we find that, although quantum fluctuations
renormalize interactions they do not modify the character
of the ground state found from the classical calculation, and the
cone state is favored.

\section{Interlayer Coupling}
\label{Eccentricity}

To this point, we have omitted the interlayer coupling $J^{\prime \prime}$ 
(except where it was essential, in order to obtain
a non-zero value for the interaction vertex $\Gamma_{\bf q} ({\bf k,k^{\prime}})$). 
It is relatively weak ($J^{\prime \prime}/J\approx 0.05$), though
crucial in stabilizing long-range order against thermal fluctuations.
As well as being small, it is also frustrated by DM interactions,
because the sign of the DM interactions alternates between 
layers (see Eq.~(\ref{hamiltonianDM})). The frustration
introduces distortions in the cone states, which we discuss in this section.

More specifically,
considering zero field for simplicity, the classical ground state in
the absence of interlayer coupling consists of a spin spiral
with wavevector ${+\bf Q}$ in layers with even index $n$,
and wavevector $-{\bf Q}$ in odd layers, as in Eq.~(\ref{spinstructureinzerofield}). 
By contrast, for a system with antiferromagnetic interlayer
exchange but no DM interactions, the ground state consists of
spin spirals with the same wavevector (say $+{\bf Q}$) in every layer,
and with alternating phases $\alpha$ in even and odd layers, so that
%===========================================================
\be
{\bf S_R}=
\begin{pmatrix} 0 \\  \cos ({\bf Q \cdot R}+ n \pi ) \\
\sin ({\bf Q \cdot R }+ n \pi )
\end{pmatrix}.
\ee
%===========================================================
With both interlayer exchange and DM interactions, their competition results in
a ground state which is a superposition of the two structures.\cite{Coldea2}
In the presence of a transverse field, spins lie on
an elliptical cone around the field direction with
%===========================================================
\ba
{\bf S_R}&=&
\begin{pmatrix} S_a({\bf R}) \\ (-1)^n S_1 \cos {\bf Q \cdot R}+
(-1)^n S_2 \cos ({\bf -Q \cdot R}) \\
S_1 \sin {\bf Q \cdot R} + S_2 (\sin {-\bf Q \cdot R})
\end{pmatrix} 
\nonumber \\&=&
\begin{pmatrix} S_a({\bf R}) \\ (-1)^n S_b \cos {\bf Q \cdot R} \\
S_c \sin {\bf Q \cdot R}
\end{pmatrix}
,
\label{spinparameter}
\ea
%===========================================================
where U(1) symmetry has been broken by selecting the 
$b$-component of the spin to alternate on adjacent layers.  The eccentricity 
%defined as
% the ratio of the major axis over the minor axis of the ellipse structure 
$I=({S_1+S_2})/({S_1-S_2})={S_b}/{S_c}$ of the cone is a
measure of the mixing of the two spin spirals at wavevectors $\pm {\bf Q}$. 

\begin{figure}[ht]
\begin{center}
\begin{picture}(200,150)
\put(0,0){\includegraphics[width=8cm]{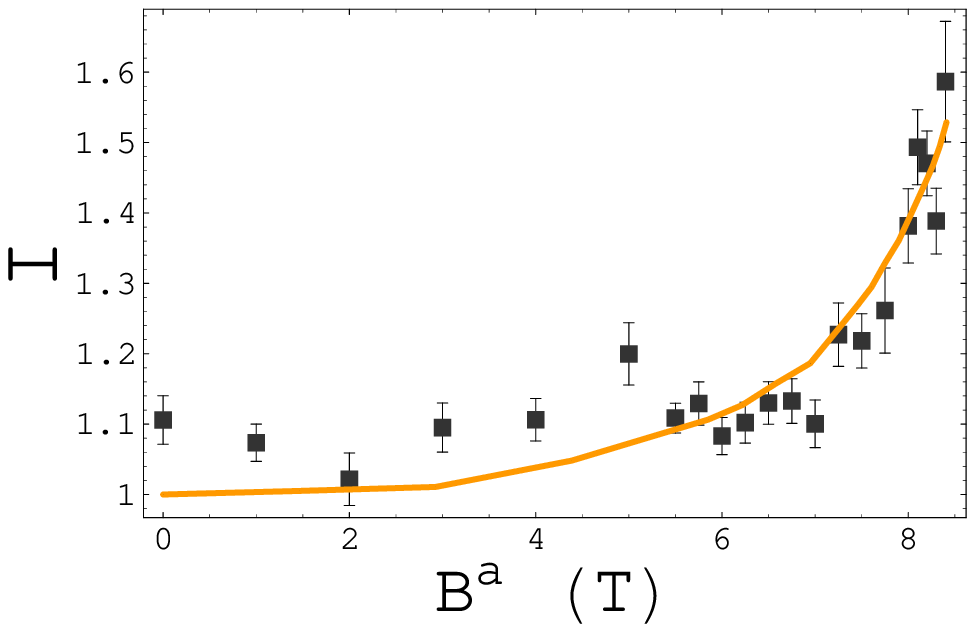}}
\put(43,78){\includegraphics[width=3.5cm]{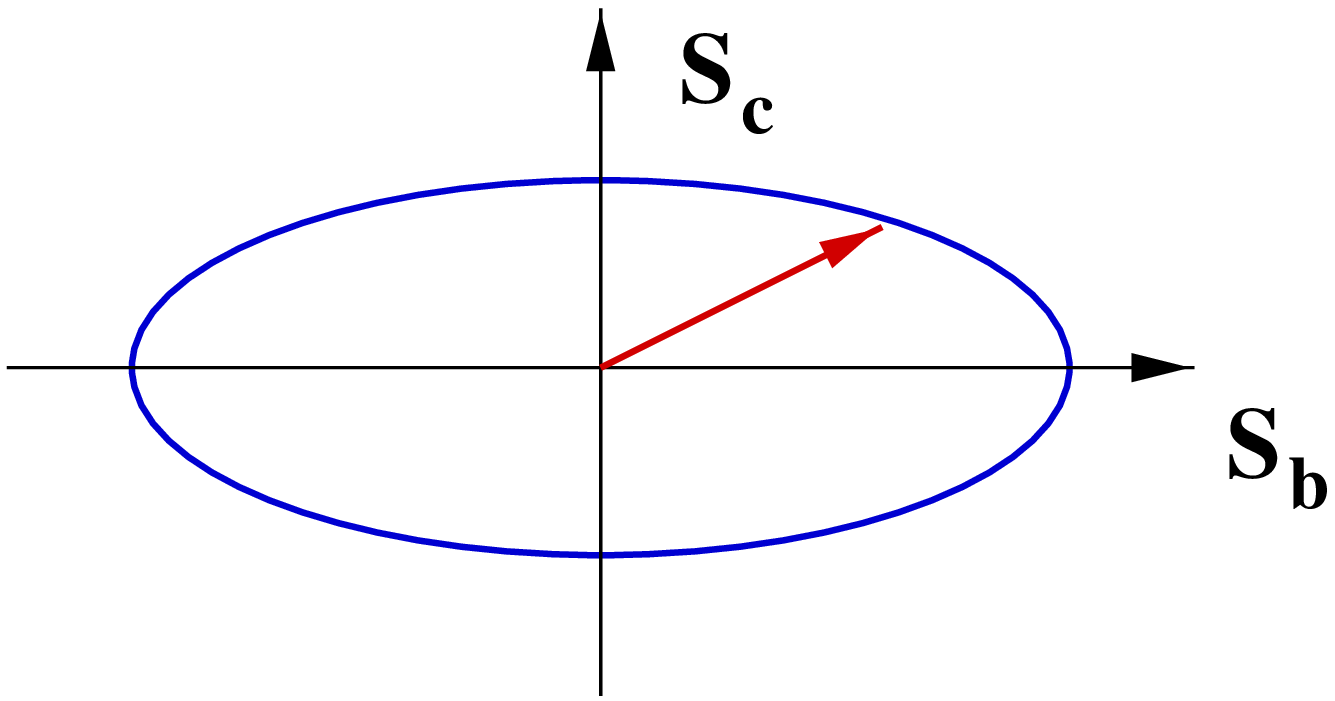}}
\end{picture}

%\vspace{0.2cm}
\caption{Eccentricity of ellipse $I=S_b/S_c$ as a function of the transverse magnetic field strength. Line: result of classical analysis. Symbols ($\Box$):
experimental data.\cite{Coldea4}}
\label{Excentricity}
\end{center}
\end{figure}

Experimentally,
this ratio can be measured by determining the relative intensity of
two magnetic Bragg peaks associated with the spin ordering,\cite{Coldea4} 
%As shown in Fig. \ref{Excentricity}, the mixing is particularly pronounced near the critical field.
and we focus on its field dependence.
%of this ratio, measured recently.\cite{Coldea4} 
The mixing between the two spin structures is observed to be
particularly strong near the critical field, where it can be calculated using
linear spinwave theory.\cite{Coldea2}
More generally, we find
the field-dependence of $I$ by
minimizing the classical energy over states which are parameterized as in
Eq.~(\ref{spinparameter}).
Results (obtained numerically) compare well with experimental data, as shown in Fig.~\ref{Excentricity}.
Mixing is small ($I\simeq 1$) in zero field, but rises rapidly near the critical 
field to reach the value $I=1.52$ at $B^a=B^a_{cr}$.

To gain insight into these results, it is useful to consider
behavior close to the critical field, and expand in powers of the small
in-plane spin components,
$S_b$ and $S_c$.
Following this procedure we obtain to quartic order the energy
%===========================================================
\ba
E&=& N \bigg[ S^2  \left(J_{\bf 0}+J^{\prime \prime}-h^a \right) 
\nonumber\\&+&  \left(J_{\bf Q} -J_{\bf
0}-J^{\prime \prime} +h^a/2 \right) \left( S_b^2+S_c^2 \right)/2
\nonumber\\+&&\!\!\!\!\!
D_{\bf Q} S_b S_c + J^{\prime \prime}\left( S^2_c -S^2_b \right)/2 
%\nonumber \\
+  \frac{J_{\bf 2 Q}- J_{\bf 0}}{32 S^2} \left(S_b^2-S_c^2
\right)^2
\nonumber\\
&+& \frac{h^a}{64 S^2} \left(3 S_b^4+3 S_c^4+ 2 S^2_b S^2_c
\right) \bigg].
\ea
This should be minimized with respect to $S_b$ and
$S_c$. It is convenient to change variables, writing $S_b=r \sin \chi$
and $S_c= r \cos \chi$, so that
\ba
E&=&N\bigg[  S^2  \left(J_{\bf 0}+J^{\prime \prime}-h^a \right)   
\nonumber\\  +&& \!\!\!\!\! \frac{r^2}{2} \left(J_{\bf Q} -J_{\bf
0}-J^{\prime \prime} +h^a/2 +
D_{\bf Q} \sin 2 \chi  + J^{\prime \prime} \cos 2 \chi \right)
\nonumber \\
&+ & r^4  \bigg( \frac{J_{\bf 2 Q}- J_{\bf 0}}{32 S^2} \cos^2 2 \chi
\nonumber\\ 
&& + \frac{h}{64 S^2} \left( 1 + 2 \cos^4 \chi +2 \sin^4 \chi   \bigg) \right)\bigg]
\label{energy functional}.
\ea

The eccentricity is then  $I=\tan \chi$. 
%\be
%I= \frac{ Max | \phi_{e {\bf Q}} e^{i {\bf Q \cdot {\bf R}}} +\phi_{o
%{\bf Q}} e^{-i {\bf Q \cdot {\bf R}}} |}{Min | \phi_{e {\bf Q}} e^{i {\bf Q \cdot {\bf R}}} +\phi_{o {\bf Q}} e^{-i {\bf Q \cdot {\bf R}}} |}
%\ee
As the critical field is approached from below, $r \rightarrow 0^+$ and
$\chi$ is determined solely by the quadratic term, 
yielding $\tan \chi =
{D_{\bf Q}}/({J^{\prime \prime}-\sqrt{(J^{\prime \prime})^2+(D_{\bf Q})^2}})=1.52$, 
as reported previously.\cite{Coldea2} 
Note that the interlayer exchange
modifies the previous estimate for the critical field (see
Sec.~\ref{Classical Analysis}) to 
$h^a_{cr}=2[J_{\bf 0}-J_{\bf Q}+J^{\prime \prime} +\sqrt{ (J^{\prime
\prime})^2+(D_{\bf Q})^2}]$, giving $B^a_{cr}=8.51 T$. 
With reducing field, $r$ increases and $I$ is determined partly
by the $\chi$ dependence of the quartic term, which is minimum at
$\chi=\pi/4 +m \pi/2$  (taking $h^a \gg J_{\bf 2 Q}-J_{\bf 0}$). 
The quartic term hence favors
$|I|=1$ and dominates as $h^a$ is reduced below $h^a_{cr}$.
% At this level, we can therefore understand the
%renormalization of the eccentricity as a competition between the
%interlayer coupling (the quadratic term) and the spin stiffness of the
%cycloid (the higher order terms) . Furthermore, since the $\chi$
%dependence of the quartic term involves $h^a$ which is much larger than $J^{\prime \prime}$ and $D_{\bf Q}$, it explains why the eccentricity ratio falls off rapidly as one moves away from the
%critical field. %In low field, the energy functional in
%Eq. \ref{energyfunctional} must be supplemented by higher order terms
%since the order parameter $r$ becomes large (i.e. of the order
%$S$). 
%The reasonable agreement between theory and experiment suggests that
%the frustration between odd and even layers is well characterized.

%%%%%%%%%%%%%%%%%%%%%%%%%%%%%%%%%%%%%%%%%%%%%%%%%%%%%%%%%%%%%%%%%%%%%%%%%%
%%%%%%%%%%%%%%%%%%%%%%%%%%%%%%%%%%%%%%%%%%%%%%%%%%%%%%%%%%%%%%%%%%%%%%%% 

\section{Conclusions}
\label{Conclusion}

In summary, we have presented a detailed investigation of ground-state
properties of an anisotropic triangular lattice antiferromagnet
with Dzyaloshinskii-Moriya interactions, focussing on behavior
in an applied magnetic field and its dependence on field direction.
We have supplemented calculations for a classical model with two
approaches to quantum fluctuations: one using linear spinwave theory,
and the other treating reversed spins in an almost polarized state
as a dilute Bose gas. We have compared our calculations with
experimental data on Cs$_2$CuCl$_4$ \cite{Coldea0,Coldea1,Coldea2,Coldea3,Coldea4,Private2}
The outcome of this comparison depends strikingly on field
direction. For a transverse field, theory is in qualitative,
and on many points quantitative agreement with experiment.
For a longitudinal field, central aspects of the low-temperature
phase diagram remain to be understood.

In more detail, for a transverse field the classical model
yields the observed incommensurate cone state with a 
field-dependent canting angle. Interlayer exchange interactions
influence this ordering in ways that are also well-described by
classical theory. However, to account for the measured field dependence of
ordering wavevector, the magnetization and the local
ordered moment, it is necessary to include the 
effect of quantum fluctuations. Linear spinwave theory
gives quite accurate results for the magnetization and qualitatively
correct behavior for the other two quantities;
the large reduction at low fields of the ordered moment below its
classical value demonstrates the importance of fluctuations.
In addition, calculations for the almost polarized system fit
observations of the ordering wavevector very well, as they should
since that aspect of the theory is asymptotically
exact.

By contrast, for a longitudinal field there are clear
differences between the phase diagram of the classical model
and experiment. In this case, classical theory yields
a distorted cycloid as the ground state at low fields,
separated by a first-order phase transition from a tilted
cone state at higher fields. Experimentally, most
of the high-field region is occupied by a third phase, 
in which no magnetic Bragg peaks have been reported,\cite{Coldea2} 
although incommensurate order has recently been observed in a narrow
field window below the saturation field.\cite{Coldea4}
Focussing on the ordering wavevector of the incommensurate phases,
spinwave theory gives only a mediocre account
of its behavior at low field, while neither spinwave theory nor
calculations for the almost polarized system can explain its value close
to saturation. In this connection, it is worth emphasizing that
the model Hamiltonian we have used must in fact omit some residual interactions
which are of importance, since it is invariant under rotations
of the magnetic field about the $a$-direction, while 
the observed phase diagram does not have exactly this symmetry.\cite{Coldea4}

For the future, the nature of the ground state in a longitudinal
field at intermediate field strengths remains an intriguing
problem, which we intend to address elsewhere.\cite{Future}

The work was supported by EPSRC under Grant GR/R83712/01 (JTC) and
Grant GR/R76714/01 (RC).

\appendix

\section{Ladder Diagram Summation}

In this appendix we show how we solve numerically the integral equation for the effective
interaction potential, $\Gamma_{\bf q} ({\bf k,k^{\prime}})$. We
recall Eq.~(\ref{Gamma})
%===========================================================
\be
\Gamma_{\bf q} = V_{\bf q} - \frac{1}{N} \sum_{\bf q^{\prime}} V_{\bf q- q^{\prime}} n_{\bf q^{\prime}} \Gamma_{\bf q^{\prime}},
\label{Gamma01}
\ee
%===========================================================
where for clarity we have omitted the variables ${\bf k, k^{\prime}}$ and introduced
%===========================================================
\be
n_{\bf q^{\prime}}= \frac{1}{\epsilon_{\bf k+q^{\prime}}+\epsilon_{\bf
k^{\prime}-q^{\prime}}-\epsilon_{\bf k}-\epsilon_{\bf k^{\prime}}}\,.
\ee
%===========================================================
The bare interaction $V_{\bf q}$ arises as the Fourier transform of an interaction
in real space, in the form $V_{\bf q}= \sum_{\bf R}
A_{\bf R} \exp (-i {\bf q} \cdot {\bf R})$. Crucially, since the
interaction is short range, only a small set of coefficients $A_{\bf R}$
are non-zero. In turn, this implies that $\Gamma_{\bf q}$
can also be expressed using a small number of Fourier coefficients, as follows.
Define the parameters
$B_{\bf R}$ through the equation $\Gamma_{\bf q}= \sum_{\bf R}
B_{\bf R} \exp (-i {\bf q} \cdot {\bf R} )$. Then from
Eq.~(\ref{Gamma01}) we obtain
%===========================================================
\be
B_{\bf R}=A_{\bf R} \left[1- \sum_{\bf R^\prime} M_{\bf R, R^{\prime}}
B_{\bf R^{\prime}} \right] ,
\label{B01}
\ee
%===========================================================
with $M_{\bf R, R^{\prime}}=\frac{1}{N} \sum_{\bf q}
n_{\bf q } e^{ -i {\bf q} \cdot ({\bf R- R^{\prime}})}$. A
simple consequence of Eq.~(\ref{B01}) is that if $A_{\bf R}=0$ for a given ${\bf R}$,
then $B_{\bf R}=0$ as well. 
%The physical interpretation of this relation is that the bare interaction term
%$A_{\bf R}$ get renormalized to yield an effective interaction $B_{\bf
%R}$ but cannot spontaneously generated.
From Eq.~(\ref{B01}) we find
%===========================================================
\be
B_{\bf R}= \sum_{R^{\prime}} \left[ A^{-1}+ M \right]^{-1}_{\bf
R^{\prime},R}\,.
\ee
%===========================================================
Since the Hamiltonian has only nearest-neighbor interactions on a stacked triangular lattice,
the matrix we must invert has $9 \times 9$ elements. These can be
evaluated numerically.

%%%%%%%%%%%%%%%%%%%%%%%%%%%%%%%%%%%%%%%%%%%%%%%%%%%%%%%%%%%%%%%%%%%%%%%%%%

\end{document}